\documentclass[prd,aps,twocolumn,pdflatex,superscriptaddress,showkeys]{revtex4-2}
\usepackage{amsmath, amsthm, amssymb,float}
\usepackage{amsfonts}
\usepackage{graphicx}
\usepackage{dcolumn}
\usepackage{bm}
\usepackage{textcomp}
\usepackage[normalem]{ulem}
\usepackage{mathrsfs}
\usepackage{color}

\usepackage{epsfig}
\usepackage{textcomp}
\usepackage{epstopdf}

\begin{document}

\title{Generation of High Order Harmonics in Vacuum  for Various Configurations of Interacting Electromagnetic Field}

\author{P. V. Sasorov}
\email{pavel.sasorov@eli-beams.eu}
\affiliation{ELI Beamlines Facility, Extreme Light Infrastructure ERIC, 252 41 Dolni Brezany,  Czech Republic}
\author{S.  V. Bulanov}
\affiliation{ELI Beamlines Facility, Extreme Light Infrastructure ERIC, 252 41 Dolni Brezany,  Czech Republic}

\date{\today}

\begin{abstract}

High-order harmonic (HOH) generation during the interaction of extremely intense electromagnetic waves in a quantum vacuum is studied within the Heisenberg-Euler formalism. 
We consider this process in the first nonvanishing order of perturbation theory. The basic expressions are derived for a general geometry, 
where the polarizations of the various sub-beams forming the focus of the electromagnetic beam are virtually identical. Explicit expressions for HOH generation are obtained 
for $4\pi$-dipole incoming wave and for two intersecting Gaussian beams. The former electromagnetic beam geometry is optimal for a given incident electromagnetic wave power, 
while the latter one is more realistic from an experimental standpoint.

\end{abstract}

\maketitle

\nopagebreak  

\section{Introduction}

High power laser  developments \cite{D2019, YP21, LI21} allow the study of nonlinear physics phenomena related to the electromagnetic field interaction with matter and
vacuum~\cite{WG1985, DG00, Mo06,Ma06,BNP94,TG09,DiP12, ZHANG20, SEI20, ROB21}.
High order harmonics (HOH) generation in vacuum caused by vacuum polarization effects in the frame of quantum electrodynamics (QED)  is one of the more important directions in these investigations \cite{AAM85, KING16, AB19, TMJ20}. This problem has attracted substantial attention 
\cite{NNR93,FG15, Ka00,VB80,P05,Lu06,Nar07,Fed07, AP14, Bohl15, FK19, SH19}
because it discloses  the dynamical properties of QED vacuum in strong electromagnetic  fields.

There is a  number of publications \cite{Ka00, P05,Lu06,Nar07,AP14,Fed07, Bohl15,Fe24} 
devoted to the detailed theoretical analysis of the harmonic generation. Nevertheless, 
the theory of this process is far from complete. Several theoretical questions are waiting for their  more thorough clarification. For example, the highest possible HOH generation rate in an optimal geometry of the incident electromagnetic (EM) wave remains unclear. One may reveal some lack of explicit analytic expressions for evaluation of the HOH intensity in more or less general but simple geometries.

We consider in this paper the $(2n+1)$-order harmonics generation in vacuum for quasi-monochromatic incident EM wave formally for any $n=1,\,2,\,\dots$. We consider the long wave limit case. It means that the wavelengths of  all the  waves taking part in the interactions are considerably larger than the Compton scattering wavelength. Thus, the theoretical consideration will be done within the framework of the Heisenberg–Euler (HE) electrodynamics~\cite{H-E, W36, ADS, BLP}. Lagrangian of the HE electrodynamics contains a perturbation to  the classic electrodynamics Lagrangian, and the perturbation can be presented as a sum of terms that can be estimated roughly as $(E/E_S)^{2m}$ $(m=2,3\dots)$, where  $E$ is a typical electric field strength in the the focus of the incident EM wave, and $E_S$ is the so called Schwinger critical field,  $E_S = m_e^2 c^3/{e\hbar} = 1.32 \times 10^{18}$~ V/m. The incident wave can be composed experimentally by several laser beams.  We present in this paper a general procedure for calculation of generated HOHs parameters. We take into account a specific form of the Heisenberg–Euler Lagrangian that shifts the HOH generation to higher order of initial laser beam intensity in comparison with naive estimation.  We show that the intensity of the $(2n+1)$-order harmonic are proportional  at least to $(E/E_S)^{4n+6}$, if we neglect high-orders corrections at $E\ll E_S$.  For this order in terms of powers of $E/E_S$ we take into account only lowest order of the QED small parameter, $\alpha=e^2/\hbar c$, the fine structure constant. The general approach can be applied for a very wide class of incident EM wave geometries. Our main goals are the explicit analytic expressions for the HOH intensities for a theoretically optimal geometry of the the incident EM wave, as well as for a geometry that is simplest from both the experimental and theoretical point of view.

There were many publications~\cite{Ba86,SSB10,StBu10,Go22}, where the optimal incident EM wave configuration was considered for observation the strong  nonlinear QED effects in vacuum. Their results indicate that the so-called  $4\pi$ dipole $in$-coming EM wave can provide maximum rate for HOH quanta production at a given $in$-coming wave power. The general theory of the HOH production will be applied in this work to obtain explicit expressions for the HOH generation for the $4\pi$ dipole $in$-coming EM wave that is most effective for the HOH generation.
We take also into account limitations arising from the possible process of the electron-positron production from the QED vacuum~\cite{Sa31,Schw51} due to the Schwinger mechanism. 
The general theory will be applied also to obtain explicit results for the HOH generation rates in a more realistic geometry of the incident EM wave: two crossing Gaussiam beams.
The rate of the HOH production in the lowest order of the laser beam intensity depends significantly on the geometry of the $in$-coming beam, and the rate drops considerably when we use weakly focused beams. However, the mechanism investigated in Ref.~\cite{Sa21}, which can be treated in the framework of the general approach, is free from this limitations and can work even in the case of colliding plane waves, though it corresponds to higher order relative to the intensity of the incident beams. We consider here a crossover between these two processes.

The paper is organized as follows. In Sec.~\ref{gen}, we present the nonlinear wave equations within the
framework of the Heisenberg–Euler electrodynamics. Sec.~\ref{2DP} is devoted to derivation 
of suitable general expressions for the HOH generation using a perturbation approach.  Sec.~\ref{3d_dipole} uses thesis theory to derive explicit expressions for the case when the $in$-coming beam forms $4\pi$ dipole field in its focus. Sec.~\ref{crossing-beams} is devoted to derivation of explicit expressions for the rate of the HOH generation for two crossing Gaussian beams and to 
comparison of the present approach with the approach of Ref.~\cite{Sa21}. The discussion and the summary of the results obtained are presented in Sec.~\ref{summ}.

\section{Equations of  the Heisenberg-Euler electrodynamics}
\label{gen}

The  analysis of the interaction of  electromagnetic waves of optical range in the QED vacuum 
 is based on the Heisenberg--Euler  Lagrangian density, $\mathcal{L}_{HE}$ \cite{H-E, W36, ADS, BLP}.  
The sum of the Lagrangians, 
\begin{equation}
\mathcal{L}=\mathcal{L}_{0}+\mathcal{L}_{HE}, \label{eq:Lagrangian}
\end{equation}
describes
the electromagnetic field in the long-wavelength limit. Here
\begin{equation}
\mathcal{L}_{0}=-\frac{m_e^4}{16\pi\alpha}F_{\mu \nu}F^{\mu \nu}
\end{equation}
 is the  Lagrangian of classical electrodynamics with  the electromagnetic field tensor $F_{\mu \nu}$ 
 defined  in terms of the 4-vector potential $A^\mu=\left(A^0,\pmb{A}\right)$  as \cite{FT}
 \begin{equation}
\label{eq:Fmn}
F_{\mu \nu}=\partial_{\mu} A_{\nu}-\partial_{\mu} A_{\nu}.
\end{equation}
Here and in the following we use a Gaussian system of units with  $\hbar=c=1$, $m_e$ and $e$ are  the electron mass   and elementary electric 
charge, and $\alpha = e^2\approx 1/137$ is the fine structure constant.  The electromagnetic fields  are measured 
in the units of $m_e^2/e$ , i.e.  are normalized on the QED critical field  $E_S = m_e^2 c^3/{e\hbar} = 1.32 \times 10^{18}$~ V/m. The mentioned above long-wavelength limit implies the typical wavelengths, $\lambda$, of the field obeys the strong inequality: $\lambda \gg \lambdabar=\hbar/m_ec$.

\begin{figure}[ht]
\centering
\includegraphics[width=0.4\textwidth, bb=0 160 800 500, clip]{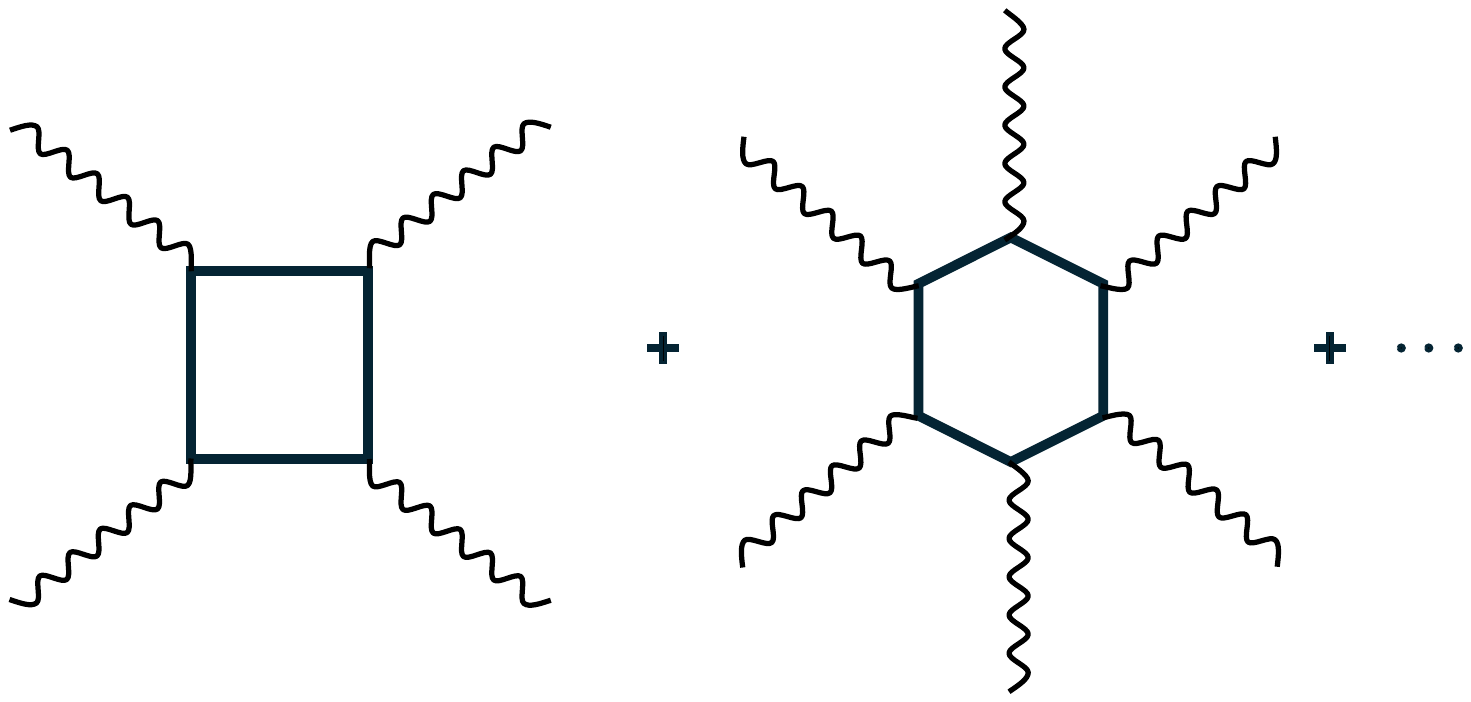}
\caption{Diagram representation of the definition~(\ref{eq:HELagr}) for $\mathcal{L}_{HE}$ as a power series of $F_{\mu\nu}$ presented below in Eq.~(\ref{Qser}).}
\label{fig00}
\end{figure}

In the Heisenberg--Euler theory, the radiative corrections are described by the $\mathcal{L}_{HE}$ 
term on the right hand side of Eq.(\ref{eq:Lagrangian}).
It can be written as \cite{BLP}
\begin{align}
\label{eq:HELagr} &
\mathcal{L}_{HE}=\frac{m_e^4} {8 \pi^2}{\cal M}( {\mathfrak e}, {\mathfrak b})= 
\frac{m_e^4}{8\pi^2}\int^{\infty}_0 \frac{\exp{(-\eta)}}{\eta^3}\times &\\
&\left[-(\eta {\mathfrak e}\cot \eta  {\mathfrak e}) (\eta {\mathfrak b}\coth \eta  {\mathfrak b}) 
+1-\frac{\eta^2}{3}({\mathfrak e}^2-{\mathfrak b}^2)\right] d \eta.\nonumber
\end{align}
The invariant fields  $ {\mathfrak e}$ and $ {\mathfrak b}$
are expressed in terms of the Poincar\'e invariants
\begin{equation}
\label{eq:FGinv}
{\mathfrak F}=\frac{1}{4}F_{\mu \nu}F^{\mu \nu}
\quad {\rm and} \quad
{\mathfrak G}={\frac{1}{4}F_{\mu \nu}\tilde F^{\mu \nu}}
\end{equation}
 as
\begin{equation}
\label{eq:abinv}
{\mathfrak e}=\sqrt{\sqrt{{\mathfrak F}^2+{\mathfrak G}^2}-{\mathfrak F}} \,\,\, {\rm and}  \,\,\, {\mathfrak b}
=\sqrt{\sqrt{{\mathfrak F}^2+{\mathfrak G}^2}+{\mathfrak F}},
\end{equation}
respectively. The dual tensor $\tilde F^{\mu \nu}$ is defined by
\begin{equation}
\tilde F^{\mu \nu}= {\frac{1}{2}}\varepsilon^{\mu \nu \rho \sigma}F_{\rho \sigma}\, ,
\end{equation}
where  $\varepsilon^{\mu \nu \rho \sigma}$ is the Levi-Civita symbol in four dimensions
{($\varepsilon^{0123}=-\varepsilon_{0123}=-1$)}.
In the 3D notations the Poincar\'e invariants are
\begin{equation}
\mathfrak{F}=f(A,A)=\frac{1}{2}\left({\bf B}^2-{\bf E}^2\right),\quad
\mathfrak{G}=g(A,A)={\bf B}\cdot{\bf E}.
\label{3d}
\end{equation}
Here and below, we consider $\mathfrak{F}$ and $\mathfrak{G}$ as diagonal values of the invariant bi-linear operator $f\left(A_1,A_2\right)$ and $g\left(A_1,A_2\right)$, respectively. They transform two electromagnetic 4-vector potential $A_1^\mu$ and $A_2^\mu$ to functions of $x$ in accordance to the following rules:
\begin{eqnarray}
f\left(A_1,A_2\right)&=&\frac{1}{2}\left(\pmb{B}_1\cdot\pmb{B}_2-\pmb{E}_1\cdot\pmb{E}_2\right)\, ;\label{N010}\\
g\left(A_1,A_2\right)&=&\frac{1}{2}\left(\pmb{E}_1\cdot\pmb{B}_2+\pmb{E}_2\cdot\pmb{B}_1\right)\, .\label{N020}
\end{eqnarray}
The Heisenberg-Euler density, $\mathcal{L}_{HE}$, presented above can be presented symbolically as a sum of the diagrams shown in Fig.~\ref{fig00} in the long-wave limit. 

As explained in Ref.~\cite{BLP},
the  Heisenberg--Euler Lagrangian in the form given by Eq.(\ref{eq:HELagr})
  should be used for obtaining an asymptotic
series over the invariant field
 $ {\mathfrak e}$ assuming its smallness.
In this limit, the function ${\cal M}( {\mathfrak e}, {\mathfrak b})$ in Eq.(\ref{eq:HELagr}) can be expanded
at small arguments as:
\begin{align}
\label{eq:M}
&{\cal M}({\mathfrak e},{\mathfrak b})=\frac{\Gamma(2)}{45}
\left[({\mathfrak e}^4+{\mathfrak b}^4)+5 {\mathfrak e}^2\,{\mathfrak b}^2\right]\nonumber \\
&-\frac{\Gamma(4)}{945}
\left[2({\mathfrak b}^6-{\mathfrak e}^6)+7{\mathfrak e}^2\,{\mathfrak b}^2({\mathfrak b}^2-
{\mathfrak e}^2)\right]\\
&+\frac{\Gamma (6)}{7\times 45^2}\left[3\left({\mathfrak e}^8+{\mathfrak b}^8\right)
+10{\mathfrak e}^2{\mathfrak b}^2\left({\mathfrak e}^4+{\mathfrak b}^4\right)-7{\mathfrak e}^4{\mathfrak b}^4\right]
+\dots \nonumber \\
\nonumber
\end{align}
Here $\Gamma(x)$ is the Euler Gamma function \cite{DLMF}.
The properties of the expansion of ${\cal M}({\mathfrak e},{\mathfrak b})$ in series of powers 
of the fields $\mathfrak e$ and  $\mathfrak b$ are discussed in Ref.~\cite{Dunne}.
The expression (\ref{eq:M}) yields for the Lagrangian $\mathcal{L}_{HE}$
in the weak field approximation  (see also Ref. \cite{HeHe})
$$ 
\mathcal{L}_{HE}=\frac{m_e^4}{8 \pi^2} {\cal M}=\frac{m_e^4}{8\times45\pi^2}\left(4 {\mathfrak F}^2
+ 7 {\mathfrak G}^2\right) 
$$
$$
 -\frac{m_e^4}{2\times315\pi^2} {\mathfrak F}
\left(  8 {\mathfrak F}^2 +13 {\mathfrak G}^2  \right)
$$
\begin{equation}
+\frac{m_e^4}{945\pi^2}\left(48\mathfrak{F}^4+88\mathfrak{F}^2\mathfrak{G}^2
+19\mathfrak{G}^4\right)+\dots
\label{eq:mathcalL}
\end{equation}
In the Lagrangian given by Eq. (\ref{eq:mathcalL}) the first, second and third terms on the right hand side 
correspond  to  four-, six- and eight-photon interactions, respectively.

In the case of  interacting electromagnetic waves, the Lagrangian given by Eq.  (\ref{eq:Lagrangian}) 
can be rewritten in terms of  $f$ and $g$,
as
\begin{equation}  \label{grangina1}
{\cal L}  = -\frac{m_e^4}{4\pi\alpha} [ f+\alpha \, Q(f,g)] =-\frac{m^4}{4\pi\alpha}\bar{\cal L}(f,g)\, .
\end{equation}
Here 
\begin{equation}  \label{barL}
\bar{\cal L}  =\bar{\cal L}_0+\bar{\cal L}_1=  f  +\alpha \, Q(f,g)
\end{equation}
is the normalized Lagrangian,
where it is assumed that  $|f|$ and $|g|$ are small compared to unity, and the function $Q(f,g)$ is
\begin{equation}\label{Qdef}
Q(f,g)=-\frac{1}{2\pi}{\cal M}\left(\sqrt{\sqrt{f^2+g^2}-f},\sqrt{\sqrt{f^2+g^2}+f}\right)\, .
\end{equation}
The function $Q(f,g)$ can be presented in the form of series
\begin{equation}\label{Qser}
Q(f,g)=\sum\limits_{m + n \geq 2} b_{m,\, n}\, f^mg^{n}\, .
\end{equation}
We skip for simplicity the arguments of the bi-linear forms $f$ and $g$ for diagonal values, when it cannot lead to confusion.

As it will be clear below in Secs.~\ref{3d_dipole} and~\ref{crossing-beams}, we will be most interested in the cases, when $|g|\ll|f|$.  For this reason, we present here only the coefficients $b_m$~\cite{Sa21}:
\begin{equation}\label{Qcoeff}
b_m=b_{m,\, 0}=\frac{2^{3(m-1)} B_{2m}}{\pi m (2m-1)(m-1)}\, .
\end{equation}
proportional to the Bernoulli numbers~\cite{DLMF}, $B_n$. 
Several leading order coefficients $b_m$ in the expansion~(\ref{Qser}) are presented in Tab.~\ref{bn}.

\begin{table}[h]
\caption{Several coefficients in the series of $Q(\zeta)$, Eq.~(\ref{Qser})}
\label{bn}
\begin{tabular}{cc|cc}
\toprule
$m$&$b_m$& $m$&$b_m$\\
\colrule
&&\\
2&$\quad-\frac{2}{45\pi}$          &8&$\quad-\frac{474087424}{26775\pi}$   \\
&&\\
3&$\quad \frac{16}{315\pi}$        &9&$\quad\frac{45997883392}{61047\pi}$        \\
&&\\
4&$\quad-\frac{64}{315\pi}$        &10&$\quad-\frac{5858972925952}{141075\pi}$   \\
&&\\
5&$\quad\frac{512}{297\pi}$         &11&$\quad\frac{20852871528448}{7245\pi}$   \\
&&\\
6&$\quad-\frac{5660672}{225225\pi}$         &12&$\quad-\frac{253794010198441984}{1036035\pi}$   \\
&&\\
7&$\quad\frac{65536}{117\pi}$         &13&$\quad\frac{5651584256049152}{225\pi}$   \\
\end{tabular}
\end{table}

The Euler-Lagrange equation corresponding to the Lagrangian~(\ref{grangina1})  gives field equations.
\begin{eqnarray}
\partial_t^2 \pmb{A}-\nabla^2 \pmb{A}
&=&4\pi \pmb{j}\left[A\right]\, ,\label{O030}\\
\partial_t^2 A^0-\nabla^2 A^0
&=&4\pi \rho\left[A\right]\, ,\label{O031}
\end{eqnarray}
where~\cite{BLP}
\begin{equation}\label{B010a}
\pmb{j}[A](\pmb{x},t)=\partial_t\frac{\partial \bar{\mathcal{L}}_1}{\partial\pmb{E}}
+\nabla\times\frac{\partial \bar{\mathcal{L}}_1}{\partial\pmb{B}}\, .
\end{equation}
and
\begin{equation}\label{B011}
\rho[A](\pmb{x},t)=-\nabla\cdot\frac{\partial \bar{\mathcal{L}}_1}{\partial\pmb{E}}
\, .
\end{equation}
We use the designations, $\pmb{j}[A]$ and $\rho[A]$, for the 3D nonlinear electric current and charge densities, respectively, to emphasize that we will express these nonlinear local functionals of $F^{\mu\nu}$ via the corresponding vector potential, $A^\mu$, in accordance to Eq.~(\ref{eq:Fmn}). 
Eqs.~(\ref{O030}) and~(\ref{O031}) are written in the form corresponding to the Lorenz gauge:
$$
\nabla\cdot \pmb{A}+\partial_t A^0=0\, .
$$

\section{Perturbation approach for the HOH generation}
\label{2DP}

\subsection{Time dependent case}

We consider in this paper a class of the problems that can be characterized as scattering problems. We assume that at $t\to-\infty$ there are several electromagnetic waves that are focused almost simultaneously at for example $t\sim 0$ at some spatial region of a typical size $L$. This asymptotic waves, which will be denoted below as $A^{in\, \mu}(\pmb{x},t)$, behaves as $\propto 1/r$ multiplied by a highly oscillating exponent, where $r$ is the large distance from this region of the focus. Hence, they obey linear wave equation:
\begin{equation}\label{O032}
\left(\partial_t^2 -\nabla^2\right) A^{in\, \mu}
=0\, .
\end{equation}
These waves are called usually as an in-coming waves.
When typical duration of these electromagnetic pulse is shorter than $L$, then we may set that the parameter $L$ is defined by this typical duration. 

Consider the parameter 
\begin{equation}\label{P010}
\eta=\alpha \omega L |F|^2\, ,
\end{equation} 
where $\omega$ is a typical frequency of the in-coming waves, and $|F|$ is a typical absolute value of $F^{ij}$ in the focus region. If $\eta\ll1$ than generation of high order harmonics can be evaluated using a perturbation theory~\cite{Sa21}. Moreover the out-going high harmonic waves can be calculated using the following 1st-order approximation~\cite{Sa21}.
\begin{equation}\label{P020}
\left(\partial_t^2 -\nabla^2\right) \pmb{A}^{out}
=4\pi\,\pmb{j}[A^{in}]\, ,
\end{equation} 
where
\begin{equation}\label{P030}
\pmb{A}^{out}=0\mbox{~~~at~~~} t\to-\infty\, ,
\end{equation} 
For the whole $\pmb{A}^{out}$ wave we should impose the condition $\pmb{A}^{out}\to \pmb{A}^{in}$ at $ t\to-\infty$. Eq.~(\ref{P020}) is actually a linear wave equation with the external nonlinear current
\begin{equation}\label{P040}
\pmb{j}^{out}=\pmb{j}[A^{in}]=\left(\partial_t\frac{\partial \bar{\mathcal{L}}_1}{\partial\pmb{E}}
+\nabla\times\frac{\partial \bar{\mathcal{L}}_1}{\partial\pmb{B}}\right)\Biggr|_{\pmb{E}^{in}\, ,\pmb{B}^{in}}\,,
\end{equation} 
which depends only on the $\pmb{A}^{in}$-field.
Thus, Eq.~(\ref{P020}) is equally a linear equation with an independent `external' current:
\begin{equation}\label{P050}
\left(\partial_t^2 -\nabla^2\right) \pmb{A}^{out}=4\pi \pmb{j}^{out}\,.
\end{equation} 

Solutions of Eq.~(\ref{P050}) at large distances from the focus (region of the beams interaction) in the direction $\pmb{n}$, obeying the condition~(\ref{P030}), can be expressed via Fourier transform $j_{\omega,\omega\pmb{n}}^{out}$ of the current
\begin{equation}\label{P052}
\pmb{j}_{\omega,\pmb{k}}^{out}=\int \pmb{j}^{out}(\pmb{x},t)e^{i\omega t-i\pmb{k}\pmb{x}}\, d\pmb{x}\, dt\,,
\end{equation}  
with $ \pmb{j}^{out}(\pmb{x},t)$ defined by Eq.~(\ref{P040}). This region of large distances is called usually as a `far zone'. It is determined by the following strong inequality: $r\gg L^2\omega$, where $L$ is typical size in the space-time localization of $ j^{out}(\pmb{x},t)$ close to the coordinate system origin: $\min\{|\pmb{x}|,|t|\}\lesssim L$.  For the case of harmonics generation we have typically $L\gtrsim \omega^{-1}$.
As a result, we have for the potential $\pmb{A}^{out}$ in the far zone ($r=|\pmb{x}|\gg L^2\omega$)
\begin{equation}\label{B570a}
\pmb{A}_{\omega}^{out}(\pmb{n} r)=\frac{e^{i\omega r}}{r}
\pmb{j}_{\omega,\, \omega \pmb{n}}^{out}\, ,
\end{equation}
and
the differential energy of radiation $d\mathcal{E}\bigr|_{\omega,\pmb{n}}$ emitted in the direction $\pmb{n}$ can be expressed as
\begin{equation}\label{P070}
d\mathcal{E}_{\pmb{n}}=
\frac{m_e^4 \left|\pmb{n}\times\pmb{j}_{\omega,\, \omega \pmb{n}}\right|^2}{4\pi^2 \alpha}\, 
\omega^2\, d\omega\, d\Phi\, .
\end{equation}
Here, $\pmb{n}=\pmb{x}/|\pmb{x}|$, $d\Phi$ is element of a solid angle in the direction $\pmb{n}$, measured in steradians, and $\omega>0$.
Similar expression can be found in Refs.~\cite{FT,Ri85}.  

The forms of Eqs.~(\ref{P040}), (\ref{P052}) and~(\ref{P070}) do not depend on the choice of the gauge. For this reason we may use an arbitrary gauge when we will use them. The transverse gauge, defined by the condition $\nabla\cdot\pmb{A}=0$, will be the most suitable for our purposes. Hence, we will imply below this gauge by default. 

We may introduce a probe out-going wave:
\begin{equation}\label{B030a}
\tilde{\pmb{A}}^{out}=\pmb{e}_{\pmb{k}}e^{-i\omega t+i\pmb{k}\pmb{x}}\, ,
\end{equation}
where $\pmb{e}_{\pmb{k}}$ is real unit vector perpendicular to $\pmb{k}$. Then, intensity of radiation of the electromagnetic wave with the polarization $\pmb{e}_{\pmb{k}}$, caused by non-linearity of the HE field equations, is determined by the following Fourier component of the nonlinear current:
\begin{widetext}
$$
\pmb{e}_{\pmb{k}}\cdot \pmb{j}_{\omega \pmb{k}}^{out}[\pmb{A}^{in}]=-\int dt\, dV \, \left\{\left(\partial_t\tilde{\pmb{A}}^{out\, *}\right)\cdot \left(\frac{\partial \bar{\mathcal{L}}_1}{\partial\pmb{E}}\right)\Bigr|_{\pmb{A}^{in}}+\left(\nabla\times\tilde{\pmb{A}}^{out\, *}\right)\cdot \left(\frac{\partial \bar{\mathcal{L}}_1}{\partial\pmb{B}}\right)\Bigr|_{\pmb{A}^{in}}\right\}=
$$
\begin{equation}\label{P060}
=-2\int dt\, dV\, \left\{\left(\frac{\partial\bar{\mathcal{L}}_1}{\partial f}\right)\biggr|_{\pmb{A}^{in}}f\left(\pmb{A}^{in},\tilde{\pmb{A}}^{out\, *}\right)+\left(\frac{\partial\bar{\mathcal{L}}_1}{\partial g}\right)\biggr|_{\pmb{A}^{in}}g\left(\pmb{A}^{in},\tilde{\pmb{A}}^{out\, *}\right)\right\}\, .
\end{equation}
\end{widetext}
This expression can be substituted into Eq.~(\ref{P070}).

Turning back to conditions of applicability of this theory, we may see that the highest value of $\eta$ takes place at the most deep focusing when $\omega L\sim 1$. We see, that even for the Schwinger field strength in the focus, when $|F|\sim 1$, the parameter $\eta$ remains small. It means that the condition $\eta\ll1$ together with the the condition $|F|\ll1$ are quite relevant for calculation of the high harmonic generation in the perturbative approach.

\subsection{Monochromatic in-coming states}
\label{mon}

It is reasonable to consider the case, when the in-coming wave $A^{in}$ is monochromatic so that it can be presented in the form:
\begin{equation}\label{P072}
\pmb{A}^{in}=\tilde{\pmb{A}}^{in}+\tilde{\pmb{A}}^{in\, *}\,;
\quad
\tilde{\pmb{A}}^{in}=e^{-i\omega t} \bar{\pmb{A}}^{in}(\pmb{x})\,\,
(\omega>0)\, ,
\end{equation}
with $\bar{\pmb{A}}^{in}(\pmb{x})$ obeying in accordance to Eq.~(\ref{O032}) the equation:
\begin{equation}\label{P072a}
\left(\nabla^2+\omega^2\right)\bar{\pmb{A}}^{in}(\pmb{x})=0\, .
\end{equation}
Then we will obtain that the out-going wave can be presented as a sum of  discrete harmonics.
\begin{equation}\label{P074}
\pmb{A}^{out}=\sum\limits_{n=1}^\infty \bar{\pmb{A}}_{(2n+1)\omega}^{out}(\pmb{x})\exp(-(2n+1)i\omega t) +\mbox{c.c.}
\end{equation}  
Combining the above equations, we obtain for the harmonic of the order $2n+1$:
\begin{equation}\label{P080}
 \left(\nabla^2+\Omega^2\right) \bar{\pmb{A}}_{\Omega}^{out}=-4\pi \bar{\pmb{j}}_{\Omega}^{out}\,,
\end{equation} 
where 
\begin{equation}\label{P081}
\Omega=(2n+1)\omega\,,\mbox{~~~and~~~}\bar{\pmb{j}}_{\Omega}^{out}=\left\langle e^{i\Omega t}\, \pmb{j}^{out}\right\rangle\,.
\end{equation} 
Here, the angle brackets $\left\langle \dots \right\rangle$ stand for averaging over the period $2\pi/\omega$. Eq.~(\ref{P080}) is written assuming the Lorenz gauge. This gauge will be used temporally in Eqs.~(\ref{P082}) and~(\ref{P084}) below to derive Eq.~(\ref{P090}), which does depend on the gauge fixation. Further we will return back to the transverse gauge, which is more suitable for the most derivations in this paper.

Solution of Eq.~(\ref{P080}) in the far zone ($r=|\pmb{x}|\gg L^2\Omega$) can be presented as
\begin{equation}\label{P082}
\bar{\pmb{A}}_\Omega^{out}(\pmb{n}\, r)=
\frac{e^{i\Omega r}}{r}\, 
\bar{\pmb{j}}_{\Omega;\,\pmb{n}\Omega}^{out}  \,,
\end{equation} 
where $\pmb{j}_{\Omega;\,\pmb{k}}^{out}$ is a spatial Fourier transform of the current $ \pmb{j}_{\Omega}^{out}(x,y)$:
\begin{equation}\label{P100}
\bar{\pmb{j}}_{\Omega;\, \pmb{k}}^{out}= \int \bar{\pmb{j}}_{\Omega}^{out}(\pmb{x})\, e^{-i\pmb{k}\cdot\pmb{x}}\, d\pmb{x}\,.
\end{equation}
We have to distinguish here $\bar{\pmb{j}}_{\Omega;\, \pmb{k}}^{out}$ from $\bar{\pmb{j}}_{\Omega,\, \pmb{k}}^{out}$ appearing in the previous section.
We have finally for the out-going $(2n+1)$ harmonic in the far zone:
\begin{equation}\label{P084}
\pmb{A}_\Omega^{out}(\pmb{n}\, r,t)=
\frac{1}{r}\,\,  \bar{\pmb{j}}_{\Omega;\,\pmb{n}\Omega}^{out}\,\,  e^{i\Omega (r-t)}
+\mbox{c.c.}\,.
\end{equation}

Eq.~(\ref{P084})  gives the following differential intensity of radiation in the direction, defined by the unit vector $\pmb{n}$:
\begin{equation}\label{P090}
d\dot{\mathcal{E}}_{\pmb{n}}=\frac{m_e^4}{2\pi\alpha}\Omega^2 \left|\pmb{n}\times\bar{\pmb{j}}_{\Omega;\, \pmb{n}\Omega}\right|^2d\Phi\, .
\end{equation}
This expression can be obtained also from Eq.~(\ref{P070}), assuming that $\pmb{j}^{out}$ is monochromatic on the time interval $-T/2<t<T/2$, where $T\gg \omega^{-1}$, and vanish sufficiently fast outside it. 

The functions $\pmb{A}^{in}$ and $\pmb{A}^{out}$ are periodic functions of time with the period $2\pi/\omega$. Using this fact, and introducing a probe out-going wave in the form:
\begin{equation}\label{P092}
\tilde{\pmb{\mathcal{A}}}^{out}(\Omega,\pmb{n})=\pmb{e}_{\Omega;\,\pmb{n}}\, e^{-i\Omega t+i\Omega \pmb{n}\pmb{x}}\, ,
\end{equation}
where $\pmb{e}_{\Omega;\,\pmb{n}}$ is unit polarization vector ($\pmb{n}\cdot \pmb{e}_{\Omega;\,\pmb{n}}=0$), we obtain a simple expression for $\bar{\pmb{j}}_{\Omega;\,\pmb{n}\Omega}$ for the monochromatic case. 
\begin{widetext}
\begin{equation}\label{P096}
\pmb{e}_{\pmb{k}}\cdot \bar{\pmb{j}}_{\Omega;
\, \pmb{k}}^{out}[\pmb{A}^{in}]=-2\int dV\, \left\langle\left(\frac{\partial\bar{\mathcal{L}}_1}{\partial f}\right)\biggr|_{\pmb{A}^{in}}f\left(\pmb{A}^{in},\tilde{\pmb{\mathcal{A}}}^{out\, *}\left(\Omega,\pmb{n}\right)\right)+\left(\frac{\partial\bar{\mathcal{L}}_1}{\partial g}\right)\biggr|_{\pmb{A}^{in}}g\left(\pmb{A}^{in},\tilde{\pmb{\mathcal{A}}}^{out\, *}\left(\Omega,\pmb{n}\right)\right)\right\rangle\, .
\end{equation}
\end{widetext} 
Here, $\pmb{A}^{in}$ is defined in Eqs.~(\ref{P072}). This expression is similar to Eq.~(\ref{P060}) and is derived by a similar way. The expression~(\ref{P096}) allows us to calculate differential intensity of generation of the harmonic of the number $2n+1$ by substituting it to Eq.~(\ref{P090}).

We may substitute the expressions~(\ref{barL}) and~(\ref{Qser}) to Eq.~(\ref{P096}) and then to Eq.~(\ref{P096}) to get a general expression for the intensity of the $(2n+1)$-harmonic. However, due to existence of several terms of the type $g^\ell f^m$ of the same order in Eq.~(\ref{Qser}), the final expression would be rather cumbersome.  For this reason, in the following consideration, we will set $g(\pmb{A}^{in},\pmb{A}^{in})=0$. Thus, our final expressions will be applicable to the case when $g(\pmb{A}^{in},\pmb{A}^{in})\ll f(\pmb{A}^{in},\pmb{A}^{in})$. We will see that quite reasonable structures of the in-coming waves can obey the latter conditions. The expression~(\ref{Qser}) contains really only even $n$. Hence, possibility of setting $g=0$ means simultaneously that we may neglect the 2nd term in Eq.~(\ref{P096}). Thus, we make the following substitution:
\begin{equation}\label{B080}
\bar{\mathcal{L}}_1\to\bar{\mathcal{L}}_1(f,g)|_{g=0}=\alpha Q(f,0)\, .
\end{equation}

Substituting the expression~(\ref{P072}) into Eq.~(\ref{P096}), applying twice the binomial theorem, and making the time averaging, we obtain:
\begin{widetext}
\begin{equation}\label{P130}
\pmb{e}_{\Omega,\,\pmb{n}}\,\bar{\pmb{j}}_{\Omega;\,\pmb{n}\Omega}^{out}=\frac{\alpha}{2\pi}
\sum\limits_{k,\, \ell \geq0}b_{n+1+\ell+2k}\frac{2^\ell (n+1+2k+\ell)!}{(n+k)!\ell! k!}H(n,\ell,k)
+\frac{\alpha}{2\pi}
\sum\limits_{k,\, \ell \geq0}b_{n+2+\ell+2k}\frac{2^\ell (n+2+2k+\ell)!}{(n+1+k)!\ell! k!}K(n,\ell,k)\,,
\end{equation}
where
\begin{equation}\label{P140}
H(n,\ell,k)=-\int f^{n+k}\left(\tilde{\pmb{A}}^{in},\tilde{\pmb{A}}^{in}\right)\,f^{\ell}\left(\tilde{\pmb{A}}^{in},\tilde{\pmb{A}}^{in\, *}\right)\, 
\,f^{k}\left(\tilde{\pmb{A}}^{in\, *},\tilde{\pmb{A}}^{in\, *}\right)f\left(\tilde{\pmb{A}}^{in},\tilde{\pmb{\mathcal{A}}}^{out\, *}\right)\, d\pmb{x}\, ,
\end{equation}
\begin{equation}\label{P150}
K(n,\ell,k)=-\int f^{n+k+1}\left(\tilde{\pmb{A}}^{in},\tilde{\pmb{A}}^{in}\right)\,f^{\ell}\left(\tilde{\pmb{A}}^{in},\tilde{\pmb{A}}^{in\, *}\right)\, 
f^{k}\left(\tilde{\pmb{A}}^{in\, *},\tilde{\pmb{A}}^{in\, *}\right)f\left(\tilde{\pmb{A}}^{in\, *},\tilde{\pmb{\mathcal{A}}}^{out\, *}\right)\, d\pmb{x}\,.
\end{equation}
\end{widetext}
We use here the following shortcut: 
$$
f^n\left(\tilde{\pmb{A}}^{in},\tilde{\pmb{A}}^{in}\right)=\left[f\left(\tilde{\pmb{A}}^{in},\tilde{\pmb{A}}^{in}\right)\right]^n\, .
$$

The lowest power of the small parameter $F$ in these sums originates from the term proportional to $H(n,0,0)$. However, $H(n,0,0)$ vanishes exactly. To explain this fact we may present $H(n,0,0)$ in the Fourier transform. Then the contribution to the integral~(\ref{P140}) comes only from the hyper-surface in the $6n+3$-dimensional $\pmb{k}$-space, where
$\pmb{k}_\Omega^{out}=\sum_{m=1}^{2n+1}\pmb{k}_m$, whereas $|\pmb{k}_\Omega^{out}|=\Omega$ and $|\pmb{k}_m|=\omega$. This conservation law can be fulfilled only when $\pmb{k}_m=\pmb{k}_\Omega^{out}/(2n+1)$ for any $m=1\,\dots , 2n+1$. However the values of $f$ entering Eq.~(\ref{P140}) vanish exactly for such combinations of the wave vectors $\pmb{k}_\Omega^{out}$ and  $\pmb{k}_m$ in accordance to the definition~(\ref{N010}). This canceling effect was considered also in Ref.~\cite{Fe24} but for the case of $n=1$ only. The authors of Ref.~\cite{KS15} did not take into account this exact canceling. They calculated $H(1,0,0)$ with an approximate procedure. This led to erroneous generation of the 3rd harmonics at a lower degree of the power of their $in$-coming wave and, hence, to considerable overestimation of the effect.

The next terms, giving lowest order of the small parameter, $F$, come from $H(n,1,0)$ and $K(n,0,0)$. We will show below that these values, $H(n,1,0)$ and $K(n,0,0)$, do not vanish in a general case. Leaving only these terms determining the leading order of intensity of high order harmonics generation, we may rewrite Eq.~(\ref{P130}) as
$$
\pmb{e}_{\Omega,\,\pmb{n}}\,\bar{\pmb{j}}_{\Omega,\pmb{n}\Omega}^{out}
=\frac{\alpha}{\pi}b_{n+2}(n+2)(n+1)H_n
$$
\begin{equation}\label{P160}
+\frac{\alpha}{2\pi}
b_{n+2}(n+2)K_n\,,
\end{equation}
where
$$
H_n=H(n,1,0)
$$
\begin{equation}\label{P170}
=-\int f^{n}\left(\tilde{\pmb{A}}^{in},\tilde{\pmb{A}}^{in}\right)\,f\left(\tilde{\pmb{A}}^{in},\tilde{\pmb{A}}^{in\, *}\right) \, f\left(\tilde{\pmb{A}}^{in},\tilde{\pmb{\mathcal{A}}}^{out\, *}\right)\, d\pmb{x}\, ,
\end{equation}
$$
K_n=K(n,0,0)
$$
\begin{equation}\label{P180}
=-\int f^{n+1}\left(\tilde{\pmb{A}}^{in},\tilde{\pmb{A}}^{in}\right)\, f\left(\tilde{\pmb{A}}^{in\, *},\tilde{\pmb{\mathcal{A}}}^{out\, *}\right)\, d\pmb{x}\,.
\end{equation}

We calculate below typical contributions to $H_n$ and $K_n$ for several typical forms of in-coming waves $\pmb{A}^{in}$ to demonstrate that they do not vanish in general case. As a result, we may say that the combination of expressions~(\ref{P090}) and~(\ref{P160})-(\ref{P180}) solves formally our problem. Thus, for the differential power of the harmonic with the number $m$, emitted in the solid angle $d\Phi$ in the direction $\pmb{n}$, we have:
\begin{widetext}
\begin{equation}\label{P180a}
d\dot{\mathcal{E}}_{\pmb{n}}=\alpha\frac{m_e^4\omega^2}{32\pi^3}m^2(m+3)^2b_{(m+3)/2}^2
\left|(m+1)H_{(m-1)/2}+K_{(m-1)/2}\right|^2\, d\Phi\,.
\end{equation}
\end{widetext}
We recall that this mechanism generates only odd harmonics. We will consider these results in more details below. Note, that the result~(\ref{P180a}) is approximate. The terms containing higher orders of $|F|$, determining the power of the $in$-coming wave, as well as higher orders of $\alpha$ are skipped from this expression. Thus, this expression is exact in this particular sense.

We should emphasize that in the case of two perfectly plane waves (of finite duration), considered in Ref.~\cite{Sa21}, the first non-vanishing effects are related to the terms $H(n,n+1,0)$ and $K(n,n,0)$. As a result, intensity of such processes typically are weaker for a focused laser beams than the processes considered above. More detailed consideration of this interesting question is placed in Sec.~\ref{comp23}.

It is possible to present $\bar{\pmb{A}}^{in}(\pmb{x})$ in the expressions~(\ref{P170}) and~(\ref{P180}) as inverse Fourier transform. It would give a momentum representation for $H_n$ and $K_n$. The momentum   representation includes integration over the $(6n + 9)$-dimensional 3-momentum space $\left\{\pmb{k}_1,\dots,\pmb{k}_{2n+3}\right\}$, with a kernel of this integral operator containing several $\delta$-functions: $\prod_{m=1}^{2n+3}\delta\left(\pmb{k}_m^2-\omega^2\right)$ and $\delta\left(\sum_{m=1}^{2n+2}\pmb{k}_m -\pmb{k}_{2n+3}-\pmb{k}_\Omega\right)$. As a result, this representation is too cumbersome, especially for $n=2,3,\dots$, and  is not convenient for our further calculations. We use below only the physical space representation for $H_n$ and $K_n$ as in Eqs.~(\ref{P170}) and~(\ref{P180}). Nevertheless, the momentum representation helps to emphasize physical meaning of our expressions. Taking in mind this momentum representation implicitly, we can say that the matrix elements~(\ref{P170}) and~(\ref{P180}) correspond to the following process expressed in terms of  4-momenta:
\begin{equation}\label{P182}
\sum_{m=1}^{2n+1} k_m^\mu+ k_{2n+2}^\mu\to k_{2n+3}^\mu + k_\Omega^\mu\,,
\end{equation} 
with all the 4-momenta belonging to mass shell of the photons, $k_\mu k^\mu=0$. The latter condition is automatically fulfilled for the matrix elements $H_n$ and $K_n$ in the momentum representation due to  Eqs.~(\ref{P072a}) and~(\ref{P092}) for $A^{in\, \mu}$ in the Lorentz gauge. Existence in Eq.~(\ref{P182}) of the `additional' photons with the momenta $ k_{2n+2}^\mu$ and $ k_{2n+3}^\mu$ provides its validity on a non-degenerate  $(4n+5)$-dimensional hyper-surface in the $(4n+8)$-dimensional manifold of the photon propagation directions $\left\{\pmb{n}_1,\dots , \pmb{n}_{2n+3}, \pmb{n}_\Omega\right\}$~\cite{Lund2}. We consider below wide enough $in$-coming beams in the space of directions of the wave 3-vectors, $\pmb{k}$. For this reason, the mentioned above property of the non-degeneracy provides that the matrix elements~(\ref{P170}) and~(\ref{P180}) for $n\geq 1$ do not vanish generally. A particular form of this statement corresponding to the case $n=0$ in our notations was previously established in Ref.~\cite{Lu06}. This almost evident conclusion will also be discussed below in Sec.~\ref{TwoBeams} for $n\geq1$.

A similar consideration but applied to the lower order matrix element $H(n,0,0)$ defined in Eq.~(\ref{P140}) leads to its exact vanishing as it is considered above. The latter matrix element corresponds to the process: $\sum_{m=1}^{2n+1} k_m^\mu\to  k_\Omega^\mu$.

The timelike  part of Eq.~(\ref{P182}) looks as the following idendity due to definition~(\ref{P081}) of $\Omega$:
\begin{equation}\label{P183}
\sum_{m=1}^{2n+1}\omega+ \omega\to \omega + \Omega\,,
\end{equation}
The first term in the right hand side of this equation, originating from the term $\pmb{k}_{2n+3}$ in Eq.~(\ref{P182}), corresponds to the beam of a macroscopic intensity, that is much larger than the intensity corresponding to the unit occupation number. From this point of view, and following Refs.~\cite{Lu06,KS15}, our classic approach treats the emission of the photons $\pmb{k}_{2n+3}$ as a stimulated emission. It will be seen also from our final results for the HOH intensities. The goal of this paper is obtaining of these expressions for the case when the HOH intensities would be high enough, including the case, when typical occupation numbers of quantum states corresponding the beam formed by the vectors $\pmb{k}_\Omega$ could be able also to be much higher than 1. For this reason, our classic approach are very reasonable for the perturbation theory for the Lagrangian~(\ref{eq:Lagrangian}). Applicability of this approach is limited by the condition that the HOH intensity is much less than the intensity of the incident wave, whereas the applicability of the quantum perturbation theory demands formally that typical occupation numbers for the HOHs are much lower than 1. Fortunately, the final results of these two approaches are mainly  the same due to the linearity of Eqs.~(\ref{P050}) and~(\ref{P080}) with respect to $\pmb{A}^{out}$, when their right hand sides are given macroscopic currents.

\section{Dipole in-coming EM field}
\label{3d_dipole}

We consider in this section an electric dipole, $E1$, {\em in}-coming field.  It corresponds to the odd photon state with $j=1$ and $m=j_z=0$~\cite{BLP,Pan05}. Here, $j$ is the total angular momentum of the photon from this EM wave, and $m$ is its $z$-component.
The full set of the photon spherical states $Ej$ and $Mj$ with given quantum numbers $j$ and $m$, corresponding to eigen states of the angular momentum operator, was considered in Refs.~\cite{BLP,Pan05} and originally in Ref.~\cite{He36}. A description of this set of the states is reproduced also in App.~\ref{EMj}. The set was constructed in Refs.~\cite{BLP,Pan05,He36} from the following representation of the exact solutions of the Maxwell equations (in the transverse gauge):
\begin{equation}\label{T410a}
\pmb{A}=\frac{e^{-i\omega t}}{(2\pi)^2}\int \delta(|\pmb{k}|-\omega)\, \pmb{a}_{\pmb{k}}
\, e^{i\pmb{k}\pmb{x}}\, d\pmb{k} +\mbox{c.c.}\, .
\end{equation}
 Here, $\omega$ is frequency of the monochromatic wave, and $\pmb{a}_{\pmb{k}}$ is tangent to the sphere $|\pmb{k}|=\omega$ in the $\pmb{k}$-space.

For the $E1$ state with $m=0$, we have in a spherical  coordinates system, $(r,\theta,\varphi)$: $B_r=B_\theta=E_\varphi=A_\varphi=0$, whereas the components $B_\varphi$, $E_\theta$, $E_r$, $A_\theta$ and $A_r$ do not vanish generally.  These components do not depend on $\varphi$. We may write for this state:
\begin{eqnarray}
B_\varphi^{in} &=&E_0\, e^{-i\omega t}\, j_1(\omega r)\sin\theta +\mbox{c.c.}
\, ,\label{T010}\\
E_\theta^{in} &=&- E_0\, e^{-i\omega t}\frac{i}{\omega r} \partial_r\left(r j_1(\omega r)\right)\sin\theta +\mbox{c.c.}
\, ,\label{T020}\\
E_r^{in} &=&E_0\, e^{-i\omega t}\,\frac{2i}{\omega r}\, j_1(\omega r)\cos\theta +\mbox{c.c.}
\, ,\label{T030}\\
A_\theta^{in} &=&E_0\, e^{-i\omega t}\frac{1}{\omega^2 r} \partial_r\left(r j_1(\omega r)\right)\sin\theta  +\mbox{c.c.}
\, ,\label{T040}\\
A_r^{in} &=&E_0\, e^{-i\omega t}\frac{2}{\omega^2 r}\, j_1(\omega r)\cos\theta  +\mbox{c.c.}
\, .\label{T050}
\end{eqnarray}
Here $j_m(.)$ is the spherical Bessel function of order $m$. 
$$
j_1(z)=\frac{\sin z -z\cos z}{z^2}\,;\quad
j_0(z)=\frac{\sin z}{z}\,;
$$
$$
j_1^\prime(z)=j_0(z)-\frac{2}{z}j_1(z)\,.
$$
The expressions~(\ref{T040}) and~(\ref{T050}) for the vector potential is written, assuming the transverse gauge fixing. We may remind that $\bar{\pmb{j}}_{\Omega;\, \pmb{n}\Omega}$ that is our intermediate goal determining the HOH generation is gauge invariant. The electromagnetic waves of types $Ej$ and $Mj$ (introduced in App.~\ref{EMj}) with $m=0$ have a specific spatial structure, so they can be called also as $TM$ and $TE$ waves, respectively.

The field~(\ref{T010})-(\ref{T050}) is normalized so that  $\pmb{E}^{in}(\pmb{x}=0,t)= (2/3)i E_0e^{-i\omega t} \pmb{e}_z +c.c.=(4/3)|E_0|\, \pmb{e}_z \sin(\omega t-\phi)$. Total power, $P_1$, of this electromagnetic wave is equal to
\begin{equation}\label{T140b}
P_1=\frac{2m_e^4|E_0|^2}{3\alpha\omega^2}\, .
\end{equation}
This structure of the in-coming electromagnetic field was considered also in Ref.~\cite{Ba86,StBu10,Gon12}. It is called in Ref.~\cite{Gon12} as $4\pi$ dipole field. This {\em in}-coming wave provides maximum value of the invariant $f$ at given laser beam intensity~\cite{Ba86}. For this electromagnetic field structure, we have that $g\left(\pmb{A}^{in},\pmb{A}^{in}\right)=0$.

We specify the expression~(\ref{P092}) for the probe out-going plane wave with frequency $\Omega=(2n+1)\omega$
$$
\bar{\pmb{\mathcal{A}}}^{out}=\pmb{e}_{\Omega;\,\pmb{n}} e^{i\Omega \pmb{n}\pmb{x}}\, .
$$
Here, $\pmb{n}=(\cos\Theta,0,\sin\Theta)$, $\pmb{e}_{\Omega;\, \pmb{n}}=(-\sin\Theta,0,\cos\Theta)$ in the Cartesian coordinate system $(x,y,z)$, associated with the spherical coordinate system, and which is used in  the expressions~(\ref{T010})-(\ref{T050}). The angle $\Theta$ ($|\Theta|\leq\pi/2$) is an angle between direction $\pmb{n}$ of the probe plane out-going wave and the plane of the dipole~(\ref{T010})-(\ref{T050}). The coefficients $H_n$ and $K_n$ depend on this angle $\Theta$. We introduce dimensionless coefficients $\bar{H}_n$ and $\bar{K}_n$ instead of $H_n$ and $K_n$ in accordance to the definitions:
\begin{equation}\label{T140a}
H_n(\Theta)=\frac{\pi(2n+1)E_0^{2n+1}|E_0|^2}{2^{n+1}\omega^2}\bar{H}_n(\Theta)
\end{equation}
\begin{equation}\label{T150a}
K_n(\Theta)=\frac{\pi(2n+1)E_0^{2n+1}|E_0|^2}{2^{n+1}\omega^2}\bar{K}_n(\Theta)
\end{equation}
Note that although the expressions in the integrands of Eqs.~(\ref{P170}) and~(\ref{P180}) are complex, nevertheless, the final results for $H_n$ and $K_n$ are real for the present in-coming field.
Explicit expressions for the coefficients $\bar{H}_n$ and $\bar{K}_n$ for the present $\pmb{A}^{in}$ are placed in App.~\ref{barHK}.

We present here final results for the case when the in-coming wave is the 3D dipole field introduced above.

Total relative power of the $(2n+1)$-harmonic can be characterized by the expression:
\begin{equation}\label{I200}
\frac{P_{2n+1}}{P_1}=\alpha^2\, a_n\, h_n\,
 |E(0)|^{4n+4}\, .
\end{equation}
Here,
\begin{equation}\label{I210}
a_n=\frac{3^{4n+5}\pi^2}{2^{10n+6}}b_{n+2}^2(n+1)^2(n+2)^2 (2n+1)^4\, ,
\end{equation}
$$
h_n=\int \left[\bar{H}_n(\Theta)+\frac{\bar{K}_n(\Theta)}{2n+2}\right]^2\, d\Phi
$$
\begin{equation}\label{I220}
=2\pi \int\limits_{-\pi/2}^{\pi/2}\left[\bar{H}_n(\Theta)+\frac{\bar{K}_n(\Theta)}{2n+2}\right]^2\, \cos\Theta\, d\Theta\, .
\end{equation}
where $|E(0)|=4|E_0|/3$ is amplitude of oscillation of electric field at $\pmb{x}=0$: $\pmb{E}^{in}(\pmb{x}=0,t)=|E(0)|\, \pmb{e}_z \sin(\omega t-\phi)$. Note, that the expression~(\ref{I220}) does not contain any parameters, besides $n$. We remind that we measure field strengths throughout the whole paper in terms of the critical Schwinger field. To obtain Eq.~(\ref{I200}) in physical units we should replace $E(0)$ in it by the ratio $E(0)/E_S$.

\begin{table*}[ht]
\caption{The coefficients $g_n$ and $h_n$ for in-coming field of type $E1$ in the 3d geometry. They enter Eq.~(\ref{I200}) for intensities of high order harmonics. They are relevant for the in-coming mode $M1$ also. The last two columns show rough estimation for maximum number of the HOH quanta that can be produced below onset of the electron-positron plasma production. }
\label{gh_3d}
\begin{tabular}{cccccc}
\toprule
&&\\
harmonic No.&$n$& $a_n$& $h_n$&~~~~~$N_{2n+1}(E1)$~~~&~~~$N_{2n+1}(M1)$~~~\\
\colrule 
\\
3&~~1~~          & ~2.26~                           & ~7.77$\times 10^{-3}$~&~$\sim2\times 10^{8}$&~$\sim2\times 10^{9}$\\
5&~~2~~        & ~88.3~                                &  ~4.13$\times 10^{-8}$~&~$\sim1$ &~$\sim 40$  \\
7&~~3~~        & ~5.36$\times 10^{3}$~& ~3.90$\times 10^{-15}$~&~$\sim2\times 10^{-10}$&~$\sim3\times 10^{-8}$\\
9&~~4~~         &  ~5.54$\times 10^{5}$~&  ~1.45$\times 10^{-22}$~&~$\sim2\times 10^{-20}$&~$\sim1\times 10^{-17}$
\end{tabular}
\end{table*}

\begin{figure}[ht]
\centering
\includegraphics[width=0.4\textwidth]{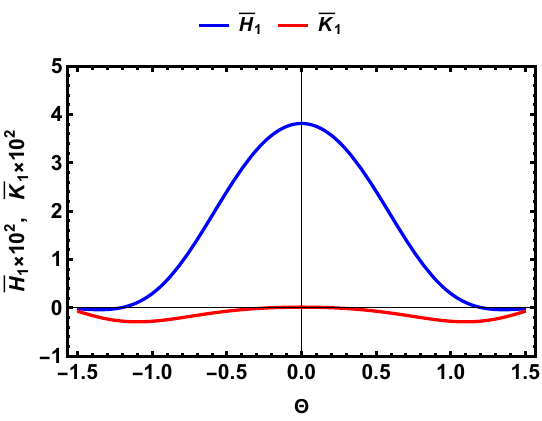}
\caption{Dependencies of
$\bar{H}_1$ (blue) and $\bar{K}_1$ (red) on the angle $\Theta$. They determine generation of the 3rd harmonics. The lines show real parts of these coefficients. For our normalization of the field $\tilde{\pmb{A}}^{in}$, the imaginary parts vanish.}
\label{fig3d1}
\end{figure}

\begin{figure}[ht]
\centering
\includegraphics[width=0.4\textwidth]{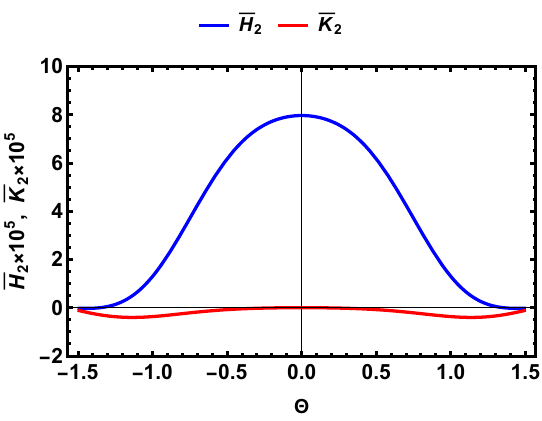}
\caption{Dependencies of  
$\bar{H}_2$ (blue) and $\bar{K}_2$ (red) on the angle $\Theta$.  They determine generation of the 5th harmonics.  The lines show real parts of these coefficients. For our normalization of the field $\tilde{\pmb{A}}^{in}$ the imaginary parts vanish.}
\label{fig3d3}
\end{figure}

\begin{figure}[ht]
\centering
\includegraphics[width=0.45\textwidth]{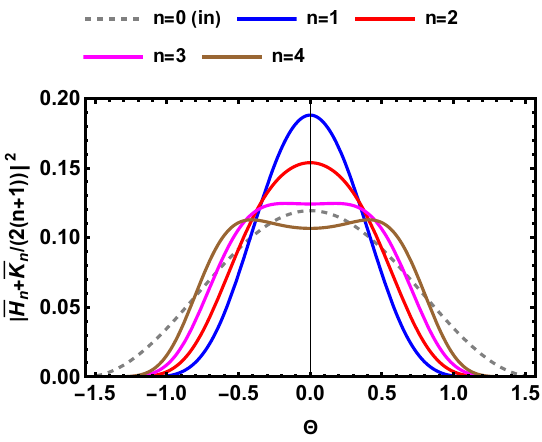}
\caption{Angular distribution of the HOH quanta. Shown is  $|\bar{H}_n+\bar{K}_n/(2n+2)|^2$, normalized on $|\bar{H}_n+\bar{K}_n/(2n+2)|^2$, integrated over the whole solid angle, $4\pi$, as functions of the angle $\Theta$ for $n=1$, 2, 3, and 4. They are proportional to angular distribution of the  harmonics of the number $m=2n+1=3,$ 5, 7, and 9. The dashed gray line shows angular distribution of the incident wave ($n=0$), normalized in the same way.}
\label{fig3d7}
\end{figure}

Tab.~\ref{gh_3d} contains numerically calculated $h_n$ as well as the numerical coefficient $a_n$ entering Eq.~(\ref{I200}). The coefficients $h_n$ are calculated numerically using the explicit expressions~(\ref{T142}) and~(\ref{T142}) from App.~\ref{barHK}  for $\bar{H}_n(\Theta) $ and $\bar{K}_n(\Theta)$. Figs.~\ref{fig3d1} and~\ref{fig3d3} show angular dependence of the coefficients $\bar{H}_n$ and $\bar{K}_n$ for $n=1$ and 2, respectively. Fig.~\ref{fig3d7} shows angular distributions of the harmonics of the numbers $(2n+1)=3$, 5, 7 and 9. They are defined by the functions $|H_n+K_n/(2n+2)|^2$, normalized by the their integrals over the whole solid angle, $4\pi$.

The magnetic dipole field, $M1$, can be obtained from the state $E1$ by means of the following substitution. The left hand side of Eq.~(\ref{T010}) should be replaced by $E_\varphi^{in}$, whereas the left hand sides of Eqs.~(\ref{T020}) and~(\ref{T030}) should be replaced by $-B_\theta^{in}$ and $-B_r^{in}$, respectively. This field generates the harmonics, which polarization vector, $\pmb{e}_{\Omega;\,\pmb{n}}$, is perpendicular to $\pmb{e}_z$, with their intensities obeying Eq.~(\ref{I200}).

Eq.~(\ref{I200}) can be rewritten equivalently in the following form.
\begin{equation}\label{I202}
\frac{3P_{2n+1}}{2P_c}=\alpha^2\, a_n\, h_n\, \left(\frac{3P_1}{2P_c}\right)^{2n+3}\, ,
\end{equation}
where
\begin{equation}\label{CC100}
P_c = \frac{m_e^4}{\alpha\omega^2}
\, ,
\end{equation}
or
\begin{equation}\label{CC102}
P_c = \frac{1}{\alpha} \left(\frac{m_ec^2}{\hbar\omega}\right)^2
\frac{m_e^2c^4}{\hbar}\approx 2\times 10^{21} \left(\frac{1~\mbox{eV}}{\hbar\omega}\right)^2
~\mbox{W}\, .
\end{equation}
We restore the physical units in this expression.

\subsection{Limitations due to electron-positron plasma creation}

We need to take into account the process of the electron-positron plasma generation from vacuum, triggered by the Schwinger mechanism~\cite{Sa31,Schw51}, and consequently amplified strongly by developing of the $\gamma e^\pm$ avalanche~\cite{DiP12,Bu06,Na04,Go22,B15,Sok10,St13,Sam21,Qu21,Mag19a,Mag19b}. The process will screen the HOH generation considered above.  We do not consider here developing of the $\gamma e^\pm$ avalanche and formation of the electron-positron plasma triggered by free electrons in the experimental chamber, when the vacuum is not sufficiently perfect.

If $\hbar\omega\sim 0.7$~eV, then probability to create one electron-positron pair in the in-coming wave per one period due to the Schwinger mechanism will be approximately equal to one half at $E(0)/E_s\approx 1/14$~\cite{Bu06,Na04,Tk24}. Such in-coming wave carries about $3\times10^{23}$ photons per period. For higher intensity of the in-coming wave, we will have intense production of the electron-positron plasma~\cite{DiP12,Go22,B15,Sok10,St13,Sam21,Qu21,Mag19a,Mag19b} that screens the generation of high order harmonics. The expression~(\ref{I200}) allows us to estimate the number of quanta of the $(2n+1)$ harmonic generated in this limiting incoming wave per its one period. Rough estimation of this number, $N_{2n+1}(E1)$, is presented in Tab.~\ref{gh_3d}. We conclude that we may hope to observe  only the 3rd and 5th harmonics if the intensity of the incoming wave below the threshold of the effective spontaneous creation of the electron-positron plasma from vacuum. Using other polarization of the initial laser beam, we can produce the $M1$ {\em in}-coming electromagnetic wave in the focus instead of $E1$. This will mitigate partially the electron-positron plasma production. It would allow us to achieve higher amplitude of the in the focus ($B(0)/E_s\simeq 1/10$)) and, hence, higher rate of the HOH generation without significant screening effect, which can be caused by electron-positron plasma production. Calculated analogous values of $N_{2n+1}(M1)$ are also shown in Tab.~\ref{gh_3d}.  Nevertheless, this limitation due to electron-positron plasma production  remains quite restrictive.

\section{Two crossing paraxial EM beams}
\label{crossing-beams}

We extend here results of Sec.~\ref{3d_dipole} to a more realistic structure of the in-coming wave. We consider here two identical Gaussian focused beams with almost parallel polarizations but intersecting at a right angle. We show that we are forced to take into account subleading corrections to pure paraxial approximation for field structure in the Gaussian beams. For this reason we start from investigation of a single beam structure. 

\subsection{3D Gaussian beam}

We start here from the following representation of any exact monochromatic solutions of the linear Maxwell equations in a vacuum.
\begin{equation}\label{T410}
\bar{\pmb{A}}=\frac{1}{(2\pi)^2}\int \delta(|\pmb{k}|-\omega)\, \pmb{a}_{\pmb{k}}
\, e^{i\pmb{k}\pmb{x}}\, d\pmb{k}\, .
\end{equation} 
Here, $\omega$ is frequency of the monochromatic wave, $\pmb{a}_{\pmb{k}}$ is tangent to the sphere $|\pmb{k}|=\omega$ in the $\pmb{k}$-space. This vector potential is written assuming the transverse gauge, $\nabla\cdot\pmb{A}=0$. Any choice of the gauge fixing is allowed for the calculation of $\bar{\pmb{j}}_{\Omega;\, \pmb{n}\Omega}$ because of its gauge invariance.

Setting $\pmb{a}_{\pmb{k}}\parallel (\pmb{e}_z\times\pmb{n})$, where $\pmb{n}=\pmb{k}/|\pmb{k}|$, we may specify $\pmb{a}_{\pmb{k}}$ by the  following way:
 $$
\pmb{a}_{\pmb{k}}=A_0\frac{ W^2}{2\pi}\left[\pmb{e}_y\cos\zeta\cos\varphi-\pmb{e}_x\cos\zeta\sin\varphi\right]
$$
\begin{equation}\label{T424}
\times\exp\left[-\frac{ W^2}{2}\left(\cos^2\zeta\sin^2\varphi+\sin^2\zeta\right)\right]\, ,
\end{equation}
where $(k,\theta=\pi/2-\zeta,\varphi)$ are spherical coordinates in the $\pmb{k}$-space, and the dimensionless parameter $W$ is a positive constant. It describes a degree of the beam converging, and defines the waist, $w_e=W/\omega$, of the Gaussian beam, when $W\gg1$. It will be considered below in more details. The exponent multiplier at $ W\gg 1$ provides a Gaussian structure of the beam. An arbitrary for the linear wave factor in front of the expression~(\ref{T424}) is chosen for our further convenience.  Thus, 
$$
\bar{\pmb{A}}(\pmb{x})=A_0  \frac{ W^2}{2\pi}
\int d\varphi\, \cos\zeta\, d\zeta
$$
$$
\times\left[\pmb{e}_y\cos\zeta\cos\varphi-\pmb{e}_x\cos\zeta\sin\varphi\right]
$$
$$
\times\exp\Bigl[i\omega\left(x\cos\zeta\cos\varphi+y\cos\zeta\sin\varphi+z\sin\zeta\right)
$$
\begin{equation}\label{T420}
- W^2\left(\cos^2\zeta\sin^2\varphi+\sin^2\zeta\right)/2\Bigr]\, .
\end{equation}
This expression gives an exact representation of the in-coming field that can be considered at $ W\to\infty$ as a model for the 3d Gaussian beam, propagating close to the $x$-axis in its positive direction. The constant complex amplitude $A_0$ describes an intensity and a phase of the wave. The expansion of Eq.~(\ref{T420}) at $ W\to\infty$ is considered in details in App.~\ref{3dGauss}. Another way of obtaining subleading corrections was considered in Refs.~\cite{La75,Da79}. Our approach differs from these papers by starting from an exact analytic expression for the solution of the Maxwell equations.

We need a couple of the first terms of the expansion of Eq.~(\ref{T420}) at $ W\gg 1$. As a result,
we can present the Gaussian beam in the following form:
\begin{equation}\label{T452a}
\left(\bar{\pmb{A}}\left(\pmb{x}\right)\right)_y=A_0\, e^{i\omega x}\, G\,\left[1+R_1+\dots\right]\, ,
\end{equation}
\begin{equation}\label{T460a}
\left(\bar{\pmb{A}}\left(\pmb{x}\right)\right)_x=A_0\, e^{i\omega x}\, G\, \left[R_2+\dots\right]\,,
\end{equation}
\begin{equation}\label{T460c}
\left(\bar{\pmb{A}}\left(\pmb{x}\right)\right)_z=0\,.
\end{equation}
Here,
\begin{equation}\label{T460b}
G=\frac{ W^2}{ W^2+i\omega x}\exp\left[-\frac{\omega^2(y^2+z^2)}{2\left( W^2+i\omega x\right)}\right]\,.
\end{equation}
The dimensionless parameter of the beam $ W$ can be expressed through the waist, $w_e$, of the beam as:
$ W^2=w_e^2\omega^2=4\pi^2w_e^2/\lambda^2$. Thus, the radius $r=w_e$ at $x=0$ corresponds the beam intensity which is equal to $1/e$ of the peak intensity of the beam at $r=0$. We assume as usually $ W\gg1$. In the latter case, 
$R_1$ and $R_2$ are of the order of ${\cal O}( W^{-2})$ and ${\cal O}( W^{-1})$, respectively, in the region, where $\omega\sqrt{y^2+z^2}\sim\mathcal{O}( W)$ and $\omega|x|\sim\mathcal{O}( W^2)$  at $ W\to\infty$.
Explicit expressions for $R_1$ and $R_2$ can be found in App.~\ref{3dGauss}. 
See Eqs.~(\ref{T450}) and~(\ref{T462}). The beam~(\ref{T452a})-(\ref{T460b}) has a circular cross section at $W\gg1$. Typical length of the waist in this limit is $W$ times larger than its typical width, $w_e$.
Setting $R_1=R_2=0$ at $ W\gg1$, we obtain a typical expression for the Gaussian beam purely in the paraxial approximation~\cite{La75,Da79}.  
Full time-dependent wave is defined as previously: $\tilde{\pmb{A}}=e^{-i\omega t} \bar{\pmb{A}}$ (assuming $\omega>0$). We have the following expression for the total power, $P$, of this beam.
\begin{equation}\label{T426}
P=\frac{1}{2}\frac{m_e^4}{\alpha} W^2\left|A_0\right|^2\, .
\end{equation}

The expressions~(\ref{T452a})-(\ref{T460b}) give that the values of $f(\tilde{\pmb{A}},\tilde{\pmb{A}})$ and $f(\tilde{\pmb{A}},\tilde{\pmb{A}}^*)$ can be estimated as
\begin{equation}\label{T480a}
|f|\sim\frac{\omega^2}{ W^2}\left|A_0\right|^2\, ,
\end{equation}
when $ W\to\infty$, while $\omega\sqrt{y^2+z^2}\sim\mathcal{O}( W)$ and $\omega|x|\sim\mathcal{O}( W^2)$.  Outside the latter region, the value of $f$ tends quickly to 0. Estimations of $g(\tilde{\pmb{A}},\tilde{\pmb{A}})$ and $g(\tilde{\pmb{A}},\tilde{\pmb{A}}^{*})$ at the same conditions give
\begin{equation}\label{T470a}
|g|\sim\frac{\omega^2}{ W^4}\left|A_0\right|^2\, .
\end{equation}

\subsection{Generation of the high-order harmonics by two intersecting Gaussian beams}
\label{TwoBeams}

\begin{figure}[ht]
\centering
\includegraphics[width=0.4\textwidth]{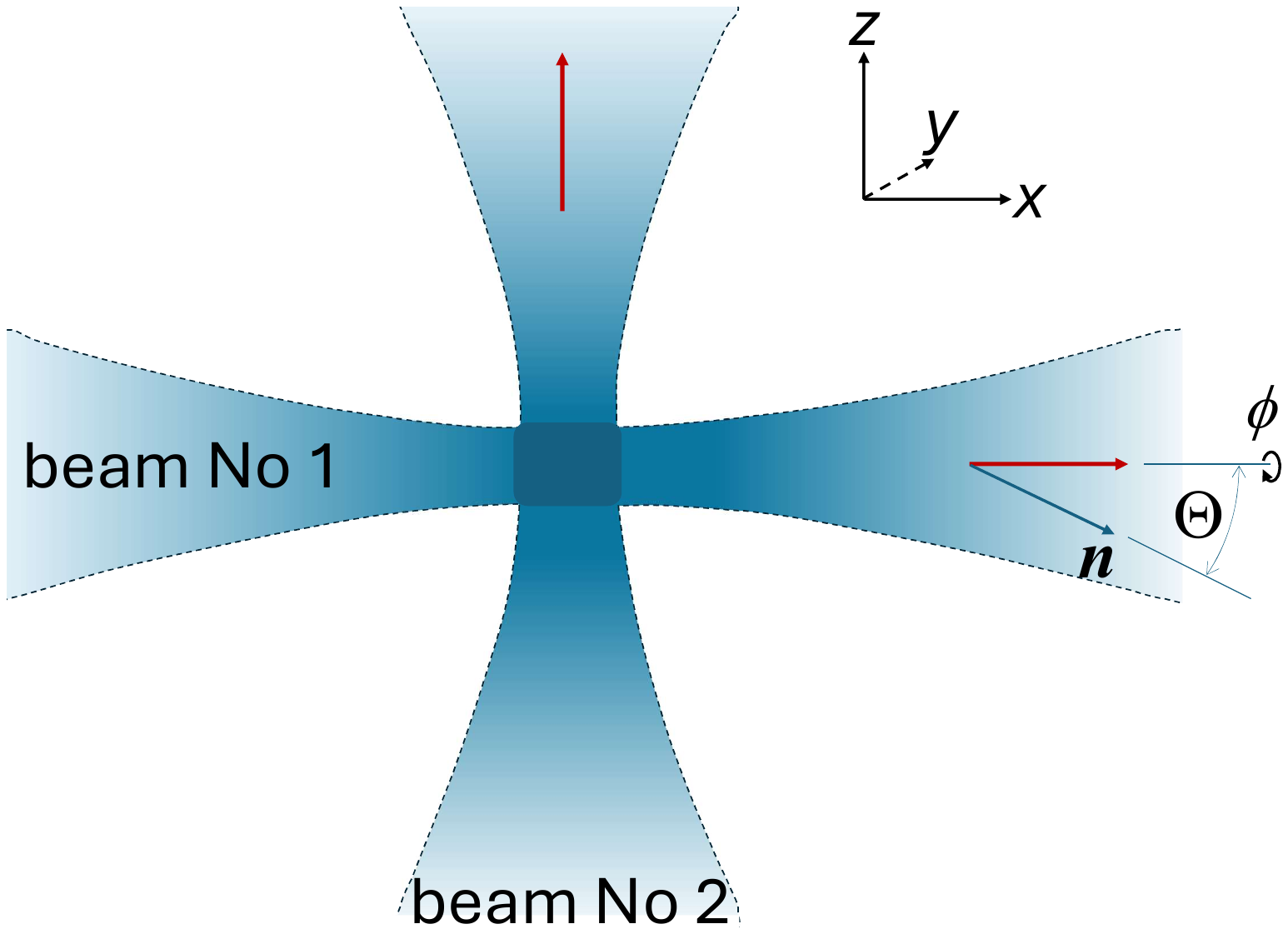}
\caption{Sketch of the two intersection Gaussian beams considered in Sec~\ref{TwoBeams}. The coordinate system used for the definitions of the incident waves structure is shown in the upper right corner.}
\label{fig7}
\end{figure}

We consider in this section generation of high order harmonics by the in-coming field combined by two Gaussian beams intersecting at a right angle.
The 1st wave corresponds as previously to the Gaussian laser beam, propagating in the $x$-direction, and polarized mainly in the $y$-direction. The 2nd wave corresponds to the beam propagating in $z$-direction with the same polarization. See Fig.~\ref{fig7} for the clarification. Results of Ref.~\cite{Sa21}, where different polarizations of two colliding beams were considered, allows us to assume that the `parallel' polarizations of the two beams is optimal for the HOH generation. Our above estimations of $|g|$, Eq.~(\ref{T470a}), say that we are able to estimate the intensities of the harmonics, using our Eqs.~(\ref{P160})-(\ref{P180a}). This approximation gives relative uncertainty in high order harmonics intensities of the order of $1/ W^4$.

The $in$-coming wave combined from two Gaussian beams of the same polarizations, intersecting at the right angle can be presented as:
\begin{equation}\label{CC510}
\bar{\pmb{A}}^{in} = \bar{\pmb{A}}_1+\bar{\pmb{A}}_2\, ,
\end{equation}
where
\begin{equation}\label{CC520}
\bar{\pmb{A}}_1(\pmb{x})=\left(\bar{A}_x(x,y,z),\bar{A}_{y}(x,y,z),0\right)\, ,
\end{equation}
\begin{equation}\label{CC530}
\bar{\pmb{A}}_2(\pmb{x})=\left(0,\bar{A}_{y}(z,y,-x),\bar{A}_{x}(z,y,-x)\right) \, ,
\end{equation}
and $\bar{\pmb{A}}$ is defined by Eqs.~(\ref{T452a})-(\ref{T460c}). 
To have an opportunity considering the beams of different intensities we will use $A_{1,\,0}$ instead of $A_0$ in the expression for $\bar{\pmb{A}}_1$ and  $A_{2,\,0}$ instead of $A_0$ in the expression for $\bar{\pmb{A}}_2$. We also specify the expression for $\pmb{\mathcal{A}}^{out}$:
\begin{equation}\label{CC270a}
\bar{\pmb{\mathcal{A}}}^{out}=\pmb{e}_{\pmb{n}} e^{i\Omega\pmb{n}\pmb{x}}\, ,
\end{equation}
where
\begin{equation}\label{CC272a}
\pmb{n}=(n_x,n_y,n_z)=(\cos\Theta,\,\sin\Theta\, \cos{\phi},\, \sin\Theta\, \sin{\phi})
\end{equation}
is the direction of the HOH propagation, and
$$
\pmb{e}_{\pmb{n}}= \pmb{n}\times\pmb{e}_{\pmb{n}}^\prime\,, 
\mbox{~~~and~~~}
\pmb{e}_{\pmb{n}}^\prime
=\frac{\pmb{n}\times\pmb{e}_y}{|\pmb{n}\times\pmb{e}_y|}\, .
$$
The angles $\Theta$ and $\varphi$ are the spherical angles associated in a standard way with the Cartesian coordinate system $(y,z,x)$. They are shown  for more clarity in Fig.~\ref{fig7}. We will use as previously that
$$
\tilde{\pmb{A}}=e^{-i\omega t} \bar{\pmb{A}}^{in}(\pmb{x})\,\,
(\omega>0)\, .
$$

There is a symmetry $x\leftrightarrow z$ in this in-coming field. When $ W\to\infty$, this means that the harmonics will be emitted in two well separated directions. One of them is directed almost along $\pmb{e}_x$, the other one is directed almost along $\pmb{e}_z$. Interesting, for example, in the HOH emitted close to the direction of the 1st beam, we may present the matrix element $H_n$ for the $2n+1$ harmonics in the following form.
\begin{equation}\label{T538}
H_n= H_n^{(2)}+H_n^{(4)}+\dots\, ,
\end{equation}
where
$$
H_n^{(2)}=-2n\int  f^{n-1}(\tilde{\pmb{A}}_1,\tilde{\pmb{A}}_1)\, f(\tilde{\pmb{A}}_1,\tilde{\pmb{A}}_2)
$$
\begin{equation}\label{T540}
\times f(\tilde{\pmb{A}}_1,\tilde{\pmb{A}}_2^*)\,
 f(\tilde{\pmb{A}}_1,\tilde{\pmb{\cal A}}^{out\, *})\, d\pmb{x}\, ,
\end{equation}
\begin{equation}\label{T560}
H_n^{(4)}=-\int f^n(\tilde{\pmb{A}}_1,\tilde{\pmb{A}}_1)\, f(\tilde{\pmb{A}}_1,\tilde{\pmb{A}}_2^*)\, 
 f(\tilde{\pmb{A}}_2,\tilde{\pmb{\cal A}}^{out\, *})\, d\pmb{x}\, .
\end{equation}
The residual term in Eq.~(\ref{T538}), designated there by dots, contains next order correction regarding $ W^{-2}\ll1$, for example the term:
$$
-\int f^n(\tilde{\pmb{A}}_1,\tilde{\pmb{A}}_1)\, f(\tilde{\pmb{A}}_1,\tilde{\pmb{A}}_1^*)\, 
 f(\tilde{\pmb{A}}_1,\tilde{\pmb{\cal A}}^{out\, *})\, d\pmb{x}\, ,
$$
We take into account only the contributions containing one pair of the terms $\pmb{A}_2$ and $\pmb{A}_2^*$ entering different $f$-forms. Other combinations of $\pmb{A}_1$, $\pmb{A}_1^*$, $\pmb{A}_2$ and $\pmb{A}_2^*$ will give exponentially small contributions to $H_n$ due to conservation of total 4-momentum of photons before and after the interaction.
Analogously, we may present $K_n$ in the following form:
$$
K_n=-2(n+1)\int f^n(\tilde{\pmb{A}}_1,\tilde{\pmb{A}}_1)
$$
\begin{equation}\label{T562}
 \times f(\tilde{\pmb{A}}_1,\tilde{\pmb{A}}_2)\,
 f(\tilde{\pmb{A}}_2^*,\tilde{\pmb{\cal A}}^{out\, *})\, d\pmb{x}+\dots\, .
\end{equation}

The functions $K_n$ and $H_n$ are determined by integrals, which integrands contain products determined by both $\tilde{\pmb{A}}_1$ and $\tilde{\pmb{A}}_2$. It means in accordance to Eqs.~(\ref{CC530}) and~(\ref{CC530}) that the integrands in Eqs.~(\ref{T540}),  (\ref{T560}) and~(\ref{T562}) are not exponentially small only inside the intersection region of the beams: $|\pmb{x}|\lesssim w_e= W/\omega$. This allows us to simplify considerably the expression for multipliers of the integrands. We will use this property below without additional mentioning. For example, we will make frequently without any mentioning the substitution in the factor in front the exponent in Eq.~(\ref{T460b}): 
$$
\frac{ W^2}{ W^2+i\omega x}\to 1\, .
$$
This multiplier is powered in the expressions for $H_n$ and $K_n$ in power $n$. Hence this substitutions is valid for high $n$ only when $n\ll  W$. This condition is not so restrictive, however. 

Main terms of the `off diagonal' multipliers $f(\tilde{\pmb{A}}_1,\tilde{\pmb{A}}_2)$,   $f(\tilde{\pmb{A}}_1,\tilde{\pmb{A}}_2^*)$,   $f(\tilde{\pmb{A}}_2,\tilde{\pmb{\cal A}}^{out\, *})$ and  $f(\tilde{\pmb{A}}_2^*,\tilde{\pmb{\cal A}}^{out\, *})$ in the expressions~(\ref{T540}), (\ref{T560}) and~(\ref{T562}) do not contain the small parameter $ W^{-2}$. For this reason, we can neglect the terms $R_1$ and $R_2$ in the expressions~(\ref{T452a}) and~(\ref{T460a}) for the Gaussian beam when calculating mentioned above bi-linear forms and can use the expressions:
\begin{equation}\label{T580}
\bar{A}_{1,\, y}=A_{1,\,0}\, \exp\left[i\omega x-\frac{\omega^2(y^2+z^2)}{2 W^2}\right]\,,
\end{equation}
\begin{equation}\label{T582}
\bar{A}_{1,\, x}=\bar{A}_{1,\, z}=\bar{A}_{2,\, x}=\bar{A}_{2,\, z}=0\,,
\end{equation}
\begin{equation}\label{T584}
\bar{A}_{2,\, y}=A_{2,\,0}\, \exp\left[i\omega z-\frac{\omega^2(x^2+y^2)}{2 W^2}\right]\,.
\end{equation}
Such simplification works 
in the domain $|\pmb{x}|\lesssim w_e= W/\omega$, giving main contribution to the integrals in Eqs.~(\ref{T540}), (\ref{T560}) and~(\ref{T562}). See for the illustration Fig.~\ref{fig7}. These multipliers behave as ${\cal O}(1)$ at $ W\to\infty$.
Thus, for the factors $f(\tilde{\pmb{A}}_1,\tilde{\pmb{A}}_2)$,   $f(\tilde{\pmb{A}}_1,\tilde{\pmb{A}}_2^*)$,
$f(\tilde{\pmb{A}}_2,\tilde{\pmb{\cal A}}^{out\, *})$ and  $f(\tilde{\pmb{A}}_2^*,\tilde{\pmb{\cal A}}^{out\, *})$, we have simple expressions in the region $|\pmb{x}|\lesssim w_e$:
$$
f(\tilde{\pmb{A}}_1,\tilde{\pmb{A}}_2)\approx \frac{1}{2}\tilde{\mathcal{E}}_{1,\, y}\tilde{\mathcal{E}}_{2,\, y}
$$
\begin{equation}\label{CA010}
=-\frac{A_{1,\,0}A_{2,\,0}\, \omega^2}{2}e^{i\omega(x+z-2t)}\exp\left(-\frac{x^2+2y^2+z^2}{2w_e^2}\right)\, ,
\end{equation}
$$
f(\tilde{\pmb{A}}_1,\tilde{\pmb{A}}_2^*)\approx \frac{1}{2}\tilde{\mathcal{E}}_{1,\, y}\tilde{\mathcal{E}}_{2,\, y}^*
$$
\begin{equation}\label{CA020}
=\frac{A_{1,\,0}A_{2,\,0}^*\omega^2}{2}e^{i\omega(x-z)}\exp\left(-\frac{x^2+2y^2+z^2}{2w_e^2}\right)\, ,
\end{equation}
$$
f(\tilde{\pmb{A}}_2,\tilde{\pmb{A}}^{out\, *})\approx \frac{1}{2}\tilde{\mathcal{E}}_{2,\, y}\tilde{\mathcal{E}}_{y}^{out\,*}
$$
\begin{equation}\label{CA030}
=-\frac{A_{2,\,0}\, \omega\Omega}{2}e^{i\Omega(t-\pmb{n}\pmb{x})+i\omega(z-t)}\exp\left(-\frac{x^2+y^2}{2w_e^2}\right)\, ,
\end{equation}
$$
f(\tilde{\pmb{A}}_2^*,\tilde{\pmb{A}}^{out\, *})\approx \frac{1}{2}\tilde{\mathcal{E}}_{2,\, y}^*\tilde{\mathcal{E}}_{y}^{out\,*}
$$
\begin{equation}\label{CA040}
=\frac{A_{2,\,0}^*\omega\Omega}{2}e^{i\Omega(t-\pmb{n}\pmb{x})+i\omega(t-z)}\exp\left(-\frac{x^2+y^2}{2w_e^2}\right)\, .
\end{equation}
Calculating Eq.~(\ref{CA030}) and~(\ref{CA040}), we set that the angle $\Theta$, defining in accordance to Eq.~(\ref{CC272a}) direction at which the HOH is observed, obeys the condition  $\Theta\lesssim 1/ W$. See for the illustration Fig.~\ref{fig7}. When $\Theta\gg 1/W$, the HOH intensity becomes exponentially low. These directions of the HOH propagation are out of our interest.

However, the factors $ f(\tilde{\pmb{A}}_1,\tilde{\pmb{\cal A}}^{out\, *})$ in~(\ref{T540}) and $f(\tilde{\pmb{A}}_1,\tilde{\pmb{A}}_1)$ in~(\ref{T560}) and~(\ref{T562}) tend to 0 at $ W\to\infty$. As a result they should be treated more accurately with taking into account the full expressions~(\ref{T452a})-(\ref{T460c}), and not their oversimplified forms~(\ref{T580})-(\ref{T584}).
Details of the latter calculations can be found in Secs.~\ref{a1a1} and~\ref{a1a_out} of App.~\ref{3dGauss}. See Eqs.~(\ref{CC260}) and~(\ref{CC290}). The calculations give:
$$
f\left(\tilde{\pmb{A}}_1,\tilde{\pmb{A}}_1\right)=A_{1,\,0}^2\omega^2\frac{e^{2i\omega (x-t)}}{2W^2}\exp\left[-\frac{(y^2+z^2)}{w_e^{2}}\right]
$$
\begin{equation}\label{CC260a}
\times\left[2i -(1+i)\frac{y^2+z^2}{w_e^2}+\frac{3y^2}{4w_e^2}+\dots\right]\, ,
\end{equation}
and
$$
f\left(\tilde{\pmb{A}}_1,\tilde{\pmb{\mathcal{A}}}^{out\, *}\right)=\frac{A_{1,\,0}\, \Omega\omega}{2 W^2} \exp\left[-\frac{(y^2+z^2)}{2w_e^2}\right]\, 
$$
$$ 
\times e^{i\omega (x-t)+i\Omega(t-\pmb{n}\pmb{x})}\biggl[\left(\frac{z\sin\phi}{w_e}+\frac{iy\cos\phi}{w_e}\right)\Theta  W
$$
\begin{equation}\label{CC290a}
 -i\frac{2w_e^2-y^2-z^2}{2w_e^2}
+(\Theta  W)^2\cos^2\phi\biggr]\, ,
\end{equation}
in the domain $|\pmb{x}|\lesssim  w_e$, giving a main contribution to the integrals. See for the illustration Fig.~\ref{fig7}. Here $\Theta$ and $\phi$ are angles defining direction of the out-going plane wave propagation. They are introduced in Eq.~(\ref{CC272a}) and illustrated in Fig.~\ref{fig7}.

The coefficients $H_n^{(2)}$, $H_n^{(4)}$ and $K_n$, defined in Eqs.~(\ref{T540})-(\ref{T562}), enter the expression~(\ref{P180a}) for the harmonics intensities only in the combination:
\begin{equation}\label{CC330a}
\mathcal{K}_n=(2n+2)H_n^{(2)}+(2n+2)H_n^{(4)}+K_n\, .
\end{equation}
Combining Eqs.~(\ref{CC260a})-(\ref{CA030}) with Eqs.~(\ref{T540})-(\ref{T562}), we obtain the following expressions for $\mathcal{K}_n$
at $ W\gg n$ and  $\Theta\lesssim 1/ W$:
\begin{widetext}
\begin{equation}\label{CC370a}
\mathcal{K}_n=\frac{(2n+1)(n+1)\sqrt{\pi}}{2^{n+2}}\, \frac{A_{1\,,0}^{2n+1}|A_{2\,,0}|^2\omega^{2n+1}}{ W^{2n-3}}\, \tilde{\mathcal{K}}_n\left( W\Theta\cos\phi, W\Theta\sin\phi\right)\, ,
\end{equation}
where
$$
\tilde{\mathcal{K}}_n\left(\tilde{\Theta}_\upsilon,\tilde{\Theta}_\zeta\right)=\int
\left[2i -(1+i)(\upsilon^2+\zeta^2)+\frac{3}{4}\upsilon^2\right]^{n-1}
$$
$$
\times\left\{2n\biggl[\zeta\tilde{\Theta}_\zeta+i\upsilon\tilde{\Theta}_\upsilon
 -i\left( 1-\frac{\upsilon^2+\zeta^2}{2}\right)
+\tilde{\Theta}_\upsilon^2\biggr]+3\left[2i -(1+i)(\upsilon^2+\zeta^2)+\frac{3}{4}\upsilon^2\right]\right\}
$$
\begin{equation}\label{CC380a}
\times \exp\left(-i(2n+1)\tilde{\Theta}_\upsilon \upsilon - i(2n+1)\tilde{\Theta}_\zeta \zeta-\frac{(2n+3)\upsilon^2+(2n+1)\zeta^2}{2}\right)
\, d\upsilon\,d\zeta\, ,
\end{equation}
\end{widetext}
$\tilde{\Theta}_\upsilon= W\Theta\cos\phi$ and $\tilde{\Theta}_\zeta= W\Theta\sin\phi$. Details of the derivations of Eqs.~(\ref{CC370a}) and~(\ref{CC380a}), are presented in Secs.~\ref{H_K} and~\ref{calK} of App.~\ref{3dGauss}. Calculation of the integral in Eq.~(\ref{CC380a}) comes down to several calculations of the integrals of the type: $\int_{-\infty}^\infty u^\ell \exp(au-bu^2)\, du$ for integer $\ell$. As a result, we may obtain explicit analytic expressions for $\tilde{\mathcal{K}}_n$ for any integer $n$.  Such expressions for $n=1$, 2, 3 and 4 are presented in App.~\ref{calK}. These explicit analytic expressions prove the fact that the matrix elements $H_n$ and $K_n$, introduced in Eqs.~(\ref{P170}) and~(\ref{P180})  do not generally vanish for this incident beams geometry. This is an example for the general statement considered in Sec.~\ref{mon}.

We are ready to derive the expression for the total power of the harmonic of the number $m=2n+1$ generated in a narrow cone around the axis $x$, i. e. around the direction of propagation the beam No 1 in Eq.~(\ref{CC510}). It will contain the value $h_n$:
\begin{equation}\label{CC400a}
h_n= W^2 \int \left|\tilde{\mathcal{K}}_n\right|^2\, d\Phi
=\int\limits_{-\infty}^\infty\int\limits_{-\infty}^\infty \left|\tilde{\mathcal{K}}_n(\tilde{\Theta}_\upsilon,\tilde{\Theta}_\zeta)\right|^2\, d\tilde{\Theta}_\upsilon\, \tilde{\Theta}_\zeta\,.
\end{equation}
We have calculated the values $h_n$ for $n=1$, 2, 3 and 4 analytically. However, these expressions are too complicated to be presented here. For example, for the simplest case $n=1$ we have
$$
h_1=\frac{533 713\, \pi^3}{97 200\sqrt{15}}\, .
$$
For this reason the values $h_n$ are presented in Tab.~\ref{GaussCrossa} in a rounded decimal form. Using Eqs.~(\ref{P180a}), (\ref{T538}), (\ref{CC330a}), (\ref{CC370a}) and~(\ref{CC400a}), we obtain for the power, $P_{1,\, m}$, of the harmonic of the number $m=2n+1$, emitted in a narrow cone around the axis $x$:
\begin{equation}\label{CA050}
\frac{P_{1,\, 2n+1}}{P_c}=\alpha^2\, \frac{a_{c,\, n}\, h_n}{ W^{8n+2}}
\left(\frac{P_{1,\, 1}}{P_c}\right)^{2n+1}\left(\frac{P_{2,\, 1}}{P_c}\right)^{2}\, ,
\end{equation}
where
\begin{equation}\label{CA060}
a_{c,\, n}=\frac{\pi (2n+1)^4(n+2)^2(n+1)^2}{32}\, b_{n+2}^2\, ,
\end{equation}
and $P_c$ is defined in Eq.~(\ref{CC100}).
Numerical values of the coefficients $a_{c,\, n}$ for $n=1$, 2, 3 and 4 are also presented in Tab.~\ref{GaussCrossa}.

\begin{table}[h]
\caption{Coefficients $h_n$ and $a_{c,\, n}$ entering Eq.~(\ref{CC400a}).}
\label{GaussCrossa}
\begin{tabular}{cccc}
\toprule
harmonic order~~~&$n$& $h_n$&$a_{c,\, n}$\\
\colrule
&&\\
3&1          &~~~43.96& ~~~$7.484\times 10^{-2}$  \\
&&\\
5&2        &~~~15.67& ~~~$3.696\times 10^{1}$        \\
&&\\
7&3        &~~~17.07& ~~~$2.839\times 10^{4}$   \\
&&\\
9&4 &~~~75.14& ~~~$3.710\times 10^{7}$   \\
\end{tabular}
\end{table}

\begin{figure}[ht]
\centering
\includegraphics[width=0.45\textwidth]{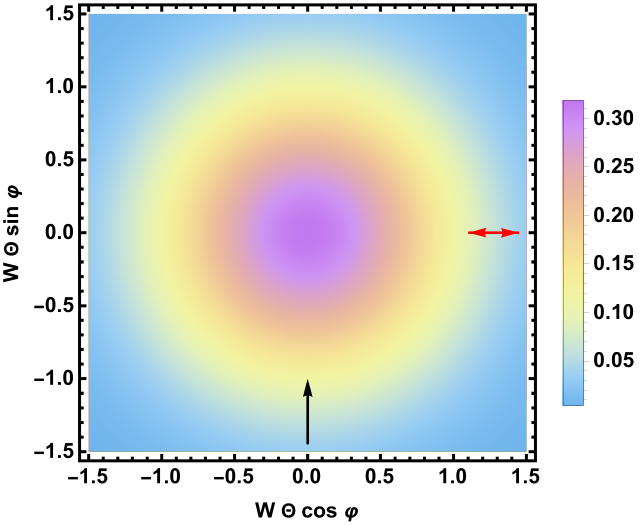}
\caption{Normalized angular distribution of the 1st (basic) harmonic of the EM beam No 1. The angles $\Theta$ and $\varphi$ are defined in Eq.~(\ref{CC272a}). The black arrow shows direction of propagation of the EM beam No. 2. The red double ended arrow shows dominant polarization of the both EM beams.}
\label{figC5}
\end{figure}

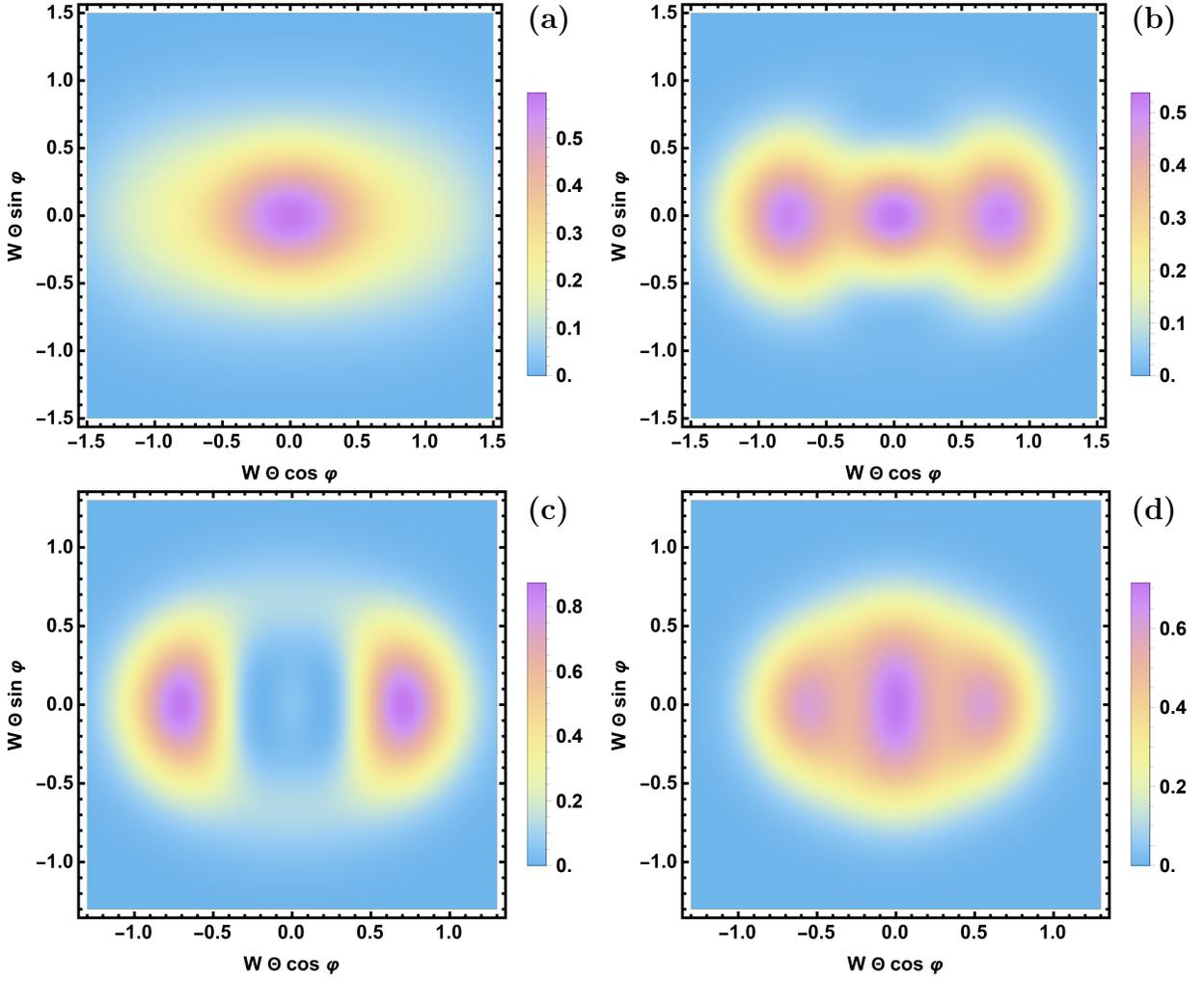
\begin{figure*}[ht]
\centering
\includegraphics[width=0.45\textwidth]{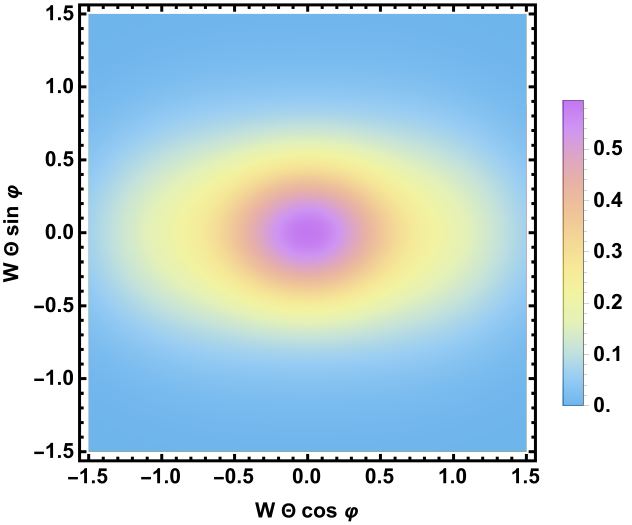}
\begin{picture}(1,180)
\put(-28,180){{\large\bf (a)}}
\end{picture}
\includegraphics[width=0.45\textwidth]{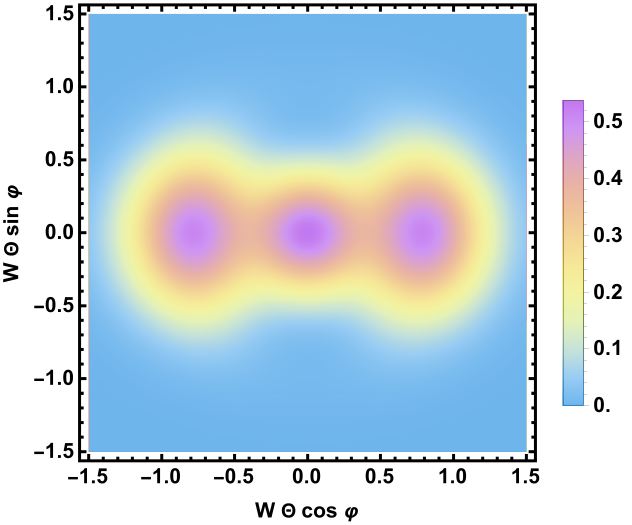}
\begin{picture}(1,180)
\put(-28,180){{\large\bf (b)}}
\end{picture}
\includegraphics[width=0.45\textwidth]{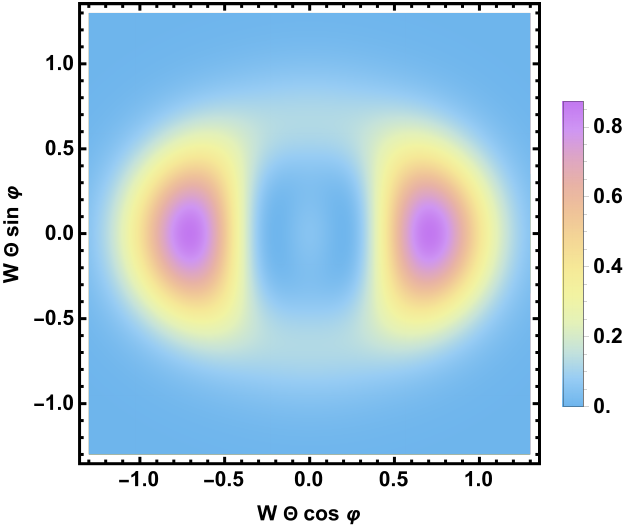}
\begin{picture}(1,180)
\put(-28,180){{\large\bf (c)}}
\end{picture}
\includegraphics[width=0.45\textwidth]{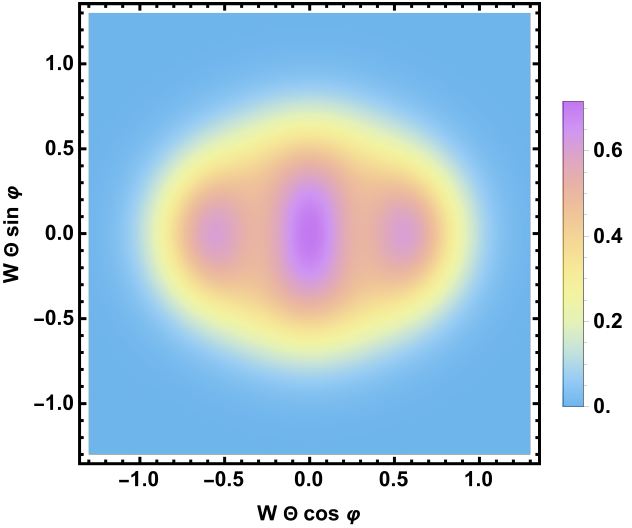}
\begin{picture}(1,180)
\put(-28,180){{\large\bf (d)}}
\end{picture}
\caption{These are normalized angular distributions of the HOH, propagating in the direction of the beam No. 1. The harmonics orders $m=2n+1=3$ , 5, 7 and 9 are shown on the panels (a), (b), (c) and (d), respectively. The angles $\Theta$ and $\varphi$ are defined in Eq.~(\ref{CC272a})}
\label{figC1}
\end{figure*}

Angular distributions of the HOH emission is determined by the expression:
\begin{equation}\label{CA064}
\frac{\left|\tilde{\mathcal{K}}_n( W\Theta\cos\phi, W\Theta\sin\phi)\right|^2}{h_n}\,.
\end{equation}
Here, $\Theta$ and $\phi$ are the angles determining the direction of the HOH emission. Their exact definition is presented in Eq.~(\ref{CC272a}). The angular distributions are presented in Fig.~\ref{figC1} for $n=1$, 2, 3 and 4, respectively. The scaled angular distributions depend on $\Theta$, $\phi$ and on $ W$. The distributions are normalized on $ W^{-2}$. Fig.~\ref{figC5} shows the analogous scaled angular distribution of the initial Gaussian beam No 1.

The HOH generation process considered in this section, whose rate is described by Eq.~(\ref{CA050}), can be described in terms of Eq.~(\ref{P182}) in the following way. The $(2n+1)$ wave 3-vectors $\pmb{k}_{1, 2, \dots, 2n+1}$ belong to the beam No~1, propagating mainly in the $x$-direction, whereas the 3-vectors $\pmb{k}_{2n+2,\, 2n+3}$ belong to the beam No~2, propagating mainly in $z$-direction. See Fig.~\ref{fig7} for the better visualization. The 3-vectors from the first group have typically small angles, $\sim 1/W$, between each other. The same is valid for the wave vectors from the second group. As a result, the matrix elements $H_n$ and $K_n$ have an additional smallness proportional to a power of $1/W$, as it was considered above. This gives the factor $1/W^{4n}$ that has appeared finally in  Eq.~(\ref{CA050}).

We see  that the amplitudes $A_{1,\, 0}$ and $A_{2,\, 0}$ enter the expression~(\ref{CC370a}) in the combination:  $A_{1,\, 0}^{2n+1}|A_{2,\, 0}|^2$.  It is interesting that this means that the phase of the harmonic is connected with the phase of the wave in the beam No 1 and does not depend on phase of the wave in the 2nd beam. Actually, the harmonic is coherent with the wave of the beam No 1 and is incoherent with the wave in the beam No 2. This situation is similar to what takes place in the works~\cite{Ko19,Ko19a,SeBu20,Sa21}.

Eq.~(\ref{CA050}) is the exact expression in the following sense. We separate the lowest (leading) order of the small parameters $E/E_S$ in an implied general expression for the intensity of the harmonic, then separate the leading order with respect to the small parameter $\alpha$, and then we separate the leading order with respect to $1/W$.    

The next order terms in the perturbation theory with respect to $\mathcal{L}_{HE}$ in Eq.~(\ref{eq:Lagrangian}) or with respect to the term $\alpha Q$ in Eq.~(\ref{barL}) will contain higher powers of $\alpha$ and, probably, lower powers of $1/W$. In this case, we should demand a relationship between $\alpha$ and $1/W$ for Eq.~(\ref{CA060}) would remain the valid leading order for the harmonic generation rate. Our preliminary estimations for the cases $n=1$, 2 and~3 show that the condition looks like: $\alpha W^2\ll 1$. We assume that this condition can be considered as a necessary condition for any $n\geq 1$. We see that Eq.~(\ref{CA050}) contains rather high powers of the small parameter $1/W$. For this reason this condition does not look as so restrictive from the practical point of view. A counterplay between the small parameters 
$E/E_S$ and $1/W$ is considered below in Sec.~\ref{comp23}.

\subsection{The limit of weakly focused EM beams}
\label{comp23}

We consider in this section a correspondence between results of the present paper and the results of the paper~\cite{Sa21}, where the interaction of two plane waves was considered. To do  this comparison we consider here again the combination~(\ref{CC510}) of two intersecting Gaussian beams as an $in$-coming wave. We consider here so high values of $ W$, entering the definition~(\ref{T452a})-(\ref{T460b}), that other terms, besides which were kept in Eq.~(\ref{P130}), begin to play an important role.

We have the general expressions~(\ref{P090}) and~(\ref{P130}) for angular distribution of emission of the HOH, when the incoming wave field has a single preferential linear polarization. Keeping the lowest order relative to the $in$-coming wave amplitude for a given harmonic order, $m=2n+1$, we obtained Eq.~(\ref{P180a}) for the angular distribution of emission of this harmonic. Eq.~(\ref{P180a}) was derived from the simplified expression~(\ref{P130}) for the non-linear current. Applying this result to the $in$-coming wave combined by two wide Gaussian beam in the form of Eq.~(\ref{CC510}), we revealed in Sec.~\ref{TwoBeams} that the electric current matrix elements from Eq.~(\ref{P160}) contain the `diagonal' values of the bi-linear form $f$: $f\left(\tilde{\pmb{A}}_1,\tilde{\pmb{A}}_1\right)$ and $f\left(\tilde{\pmb{A}}_1,\tilde{\pmb{\mathcal{A}}}^{out\, *}\right)$. Eqs.~(\ref{CC260a}) and~(\ref{CC290a}) present the expressions of these values relevant for the crossing Gaussian beams. The condition $ W=w_e/\omega \gg 1$ provides existence of the additional small factor $ W^{-2}$ in the values of the bi-linear form.  This fact leads to existence of the additional factor $ W^{-4n}$ in the intensity of the harmonic of the order $m=2n+1$ at the given wave amplitude $|A_0|$. See Eq.~(\ref{CA050}). It contains also additional, more trivial powers of $ W^{-1}$ because of the relationship~(\ref{T426}) between $E(0)$ and $P$. Turning now to the almost plane wave at $ W\to\infty$ that is the case of Ref.~\cite{Sa21}, we see that this additional factor tends to 0. A finite results of Ref.~\cite{Sa21}, obtained there for the plane waves, were obtained by keeping in the current only the `off diagonal' contributions to the bi-linear form. See, for example, Eq.~(60) of Ref.~\cite{Sa21}. We may see, however, that the leading orders (at $ W\to\infty$) of the next order contributions relative to $A_0$ (at $|A_0|\to 0$) in the sums over $\ell$ in Eq.~(\ref{P130}) at given $n$ contain lower powers of $ W^{-1}$. The terms containing $H(n,n+1,0)$ and $K(n,n,0)$ will not contain at all the factor $W^{-1}$ in their leading orders at $ W\to\infty$ the `diagonal' values of $f$ and, hence, this small factor. These terms are the lowest terms in powers of $A_0$, demonstrating this property. 

We can rewrite the expressions~(\ref{P140}) and~(\ref{P150}) for $H(n,n+1,0)$ and $K(n,n,0)$, respectively, at $ W\to\infty$ in the following forms.
$$
H(n,n+1,0)=
- 2^{2n+1} \int f^n\left(\tilde{\pmb{A}}_1,\tilde{\pmb{A}}_2\right) 
$$
\begin{equation}\label{CC070a}
\times f^{n+1}\left(\tilde{\pmb{A}}_1,\tilde{\pmb{A}}_2^*\right)
 f\left(\tilde{\pmb{A}}_2,\pmb{\cal A}^{out\, *}\right) \, dV+\dots\, .
\end{equation}
and
$$
K(n,n,0)=
- 2^{2n+1}\int f^{n+1}\left(\tilde{\pmb{A}}_1,\tilde{\pmb{A}}_2\right)
$$
\begin{equation}\label{CC072a}
\times  f^{n}\left(\tilde{\pmb{A}}_1,\tilde{\pmb{A}}_2^*\right) \,
 f\left(\tilde{\pmb{A}}_2^*,\pmb{\cal A}^{out\, *}\right) \, dV+\dots\, .
\end{equation}
We keep here only `off diagonal' values of $f(\dots)$.
The terms~(\ref{CC070a}) and~(\ref{CC072a}) enter Eq.~(\ref{P130}) only in the following combination.
\begin{equation}\label{CA070}
M_n=2H(n,n+1,0)+K(n,n,0)\, .
\end{equation}
Substituting the expressions~(\ref{CA010})-(\ref{CA040}) for $f\left(\tilde{\pmb{A}}_1,\tilde{\pmb{A}}_2\right)$, $ f\left(\tilde{\pmb{A}}_1,\tilde{\pmb{A}}_2^*\right)$, 
$f\left(\tilde{\pmb{A}}_2,\pmb{\cal A}^{out\, *}\right)$ and  $f\left(\tilde{\pmb{A}}_2^*,\pmb{\cal A}^{out\, *}\right)$, respectively, into Eqs.~(\ref{CC070a}) and~(\ref{CC072a}), we obtain:
$$
M_n=\frac{3(-1)^{n+1} (2n+1)^{2n+1}}{2^{2 n+2}}A_{1,\,0}^{2n+1} |A_{2,\, 0}|^{2n+2}\omega^{4n+4}
$$
$$
\times\int\exp\left(-\frac{(2n+2)x^2+(4n+3)y^2+(2n+1)z^2}{2w_e^2}\right)
$$
$$
\times\exp\left(i\Omega (x-\pmb{n}\pmb{x})\right)\, dx\, dy\, dz
$$
$$
=\frac{3(-1)^{n+1} (2n+1)^{2n+1}}{2^{2 n+1}}
$$
$$
\times  \sqrt{\frac{\pi^3}{(n+1)(4n+3)(2n+1)}} \,A_{1,\,0}^{2n+1} |A_{2,\, 0}|^{2n+2} W^3\omega^{4n+1}
$$
\begin{equation}\label{CA080}
\exp\left[-\frac{(2n+1)^2}{2(4n+3)}  W^2\Theta^2\left(1+\frac{2n+2}{2n+1}\sin^2\phi\right)\right]\,.
\end{equation}

Assume for a moment that the matrix elements $H(n,n+1,0)$ and $K(n,n,0)$ in the form of Eq.~(\ref{CA070}) give a dominant contribution to the electric current matrix element~(\ref{P130}). We specify below when the assumption is valid. In this case we can keep only the expression~(\ref{CA080}) for the calculation of the total power of the harmonic of the order $m=2n+1$ emitted into a narrow cone around the direction of propagation of the initial beam No 1, i.e., around the axis $x$. Using Eqs.~(\ref{P090}) and~(\ref{P130}), and integrating Eq.~(\ref{P090}) over $d\Phi$, we obtain:
\begin{equation}\label{CA090}
\frac{P_{1,\, 2n+1}}{P_c}=\alpha^2  \frac{a_{p,\, n}}{ W^{8n+2}}
\left(\frac{P_{1,\, 1}}{P_c}\right)^{2n+1}\left(\frac{P_{2,\, 1}}{P_c}\right)^{2n+2}
\end{equation}
where
\begin{equation}\label{CA100}
a_{p,\, n}=\frac{9\pi2^{2n-2} (2n+1)^{4n+2} (2n+2)!^2}{n!^2(n+1)!^2(n+1)\sqrt{(2n+1)(4n+3)}}b_{2n+2}^2\,,
\end{equation}
and $P_c$ is defined in Eq.~(\ref{CC100}). The values of the coefficients $a_{p,\, n}$ are presented in Tab.~\ref{TabPlane}. 
Relative angular distribution of this radiation  is described by the square of the last multiplier in Eq.~(\ref{CA080}):
\begin{equation}\label{CA110}
\exp\left[-\frac{(2n+1)^2}{4n+3}  W^2\Theta^2\left(1+\frac{2n+2}{2n+1}\sin^2\phi\right)\right]\, .
\end{equation}
It is normalized to 1 at its maximum, at $\Theta=0$.

\begin{table}[h]
\caption{Coefficients  $a_{p,\, n}$, defined in Eq.~(\ref{CA100}), and entering Eq.~(\ref{CA090}).}
\label{TabPlane}
\begin{tabular}{ccc}
\toprule
harmonic order~~~&$n$&$a_{p,\, n}$\\
\colrule
&&\\
3&1          & ~~~1.858 \\
&&\\
5&2        & ~~~$1.171\times 10^6$          \\
&&\\
7&3        & ~~~$2.749\times 10^{13}$   \\
&&\\
9&4 & ~~~$7.679\times 10^{21}$   \\
&&\\
11&5 & ~~~$1.417\times 10^{31}$   \\
\end{tabular}
\end{table}

The result~(\ref{CA090}) can be also obtained after some algebra from the results of Ref.~\cite{Sa21} by means of applying a proper  Lorentz transform. Nevertheless, we prefer here to derive it from our general expressions of Sec.~\ref{mon}.

Comparing now the results~(\ref{CA050}) and~(\ref{CA090}), we reveal that they depend similar on $ W$ and on $P_1$, but differently on the power of the 2nd beam, $P_2$. For higher powers, $P_2$,  the mechanism~\cite{Sa21}, considered in this sub-section, dominates over the lowest order mechanism in terms of the strength of the wave field, considered in Sec.~\ref{TwoBeams} and corresponding to the general expression~(\ref{P180a}). For the lower powers, $P_2$, the mechanism, leadig to Eq.~(\ref{CA090}) dominates. The crossover between these two mechanisms of the HOH generation takes place at
\begin{equation}\label{CA120}
P_{2,\,0}\bigr|_c= P_c \times \left(\frac{a_{c,\, n}h_n}{a_{p,\, n}}\right)^{1/2n}\, ,
\end{equation}
where the characteristic laser beam power, $P_c$, is defined in Eq.~(\ref{CC100}). The dimensionless multiplier, depending on the harmonic order, $m=2n+1$, is equal to $\approx 0.030$ for $m=3$ and to $\approx 0.037$ for $m=5$. Spontaneous electron-positron pairs production is determined by $\mathcal{E}=\sqrt{-2\min f(A^{in},A^{in})}$~\cite{BLP}. Assuming for simplicity that $P_1=P_2$, we obtain that $\mathcal{E}/E_S\approx 0.7/ W$ for $m=3$ and $\approx 0.8/ W$ for $m=5$ at the point of the crossover. We conclude both effects can act in the regime below the onset of the electron-positron pair production, when $ W\gg 1$. The latter condition is assumed to be valid when we consider the two crossing Gaussian beams. To avoid possible misunderstanding, we recall that, in the case of the 3d-dipole $in$-coming wave considered in Sec.~\ref{3d_dipole}, there is no room for the mechanism based on the matrix elements $H(n,n,0)$ and $K(n,n+1,0)$ before the onset of the electron-positron plasma production at optical frequencies, as it was considered in Sec.~\ref{3d_dipole}.

\section{Summary and Discussion}
\label{summ}

A general theory of the HOH generation obtained in the frame of the Heisenberg-Euler Electrodynamics
in vacuum is presented in this paper. The HE electrodynamics is based on existence of two small parameters: $\alpha$, the fine structure constant, and $E/E_S$, where $E$ is a typical absolute value of the electric field strength in the focus of the incident EM wave, and $E_S$ is the QED critical field. We obtain the general expression~(\ref{P180a}) for the angular distribution of the HOH intensity for the case of monochromatic $in$-coming wave. 
The case of the `same' polarization is only considered in this paper. This limitation simplifies considerably the final expressions, and takes into account an observation that typically mixed polarization leads to diminishing of the HOH generation rate~\cite{Sa21}. Our general expression~(\ref{P180a}) is an exact leading order term of a general expression in the following sense. We keep only the lowest powers of the small parameters $E/E_S$ and $\alpha$.

We apply these general expressions for the two cases: i) for the $4\pi$ dipole $in$-coming field that gives maximal efficiency for the HOH generation at given total power of the $in$-coming wave; and ii) for the case of two crossing Gaussian beams. These results are summarized in i) Eq.~(\ref{I200}) and Tab.~\ref{gh_3d}, and in ii) Eq.~(\ref{CA050}) and Tab.~\ref{GaussCrossa}, respectively. Angular distributions of the HOH radiation are presented in Figs.~\ref{fig3d7} and~\ref{figC1}, respectively. The expressions for the case (ii) contain an additional small parameter, $1/W$, that characterizes a small convergence angle of the Gaussian beams.  A counterplay of this small parameter with $E/E_S$ and $\alpha$ is considered in Secs.~\ref{TwoBeams} and~\ref{comp23}.

Power of the HOH of the order $m$, generated by the mechanism considered in this paper,  is proportional to $P^{m+2}$, where $P$ is total power of the $in$-coming wave. However, we cannot increase the power unlimitedly. At a sufficiently high power, the wave will produce electron-positron plasma~\cite{Bell,DiP12,Go22,B15,Sok10,St13,Sam21,Qu21,Mag19a,Mag19b} as an avalanche triggered by the Schwinger mechanism~\cite{Sa31,Schw51,Bu06,Na04} of the electron-positron pair production. Radiation of such a plasma will lead to screening of the observable effect of HOH generation in vacuum. Thus, total rate of the HOH generation in vacuum that can be investigated experimentally is bounded from above. We considered this effect and the results are summarized in Tab.~\ref{gh_3d} separately for the $E1$ and $M1$ mode for the $4\pi$-dipole $in$-coming wave. We see that it would be hardly to detect the harmonics with the number $m$ higher than 5 for optical lasers.

Our general expression~(\ref{P180a}), containing the matrix elements~(\ref{P170}) and~(\ref{P180}),    is obtained in the lowest order of our perturbation theory relative to $in$-coming wave power. They corresponds to the process
\begin{equation}\label{SD010}
k_1+k_2+\dots+ k_{2n+2}\to k_{2n+3}+\mathcal{K}\, ,
\end{equation}
if we describe it in terms of elementary plane waves.
Here the wave 4-vectors, $k_{i=1,2,\dots, 2n+3}$, belong to the $in$-coming wave (beams), whereas the 4-wave vector, $\mathcal{K}$, corresponds to the generated HOH beam. All $k_i$ have the same frequency $\omega$, whereas the time component of the vector $\mathcal{K}$ corresponds to the frequency $\Omega=(2n+1)\omega$ of the HOH. 

Conservation of 4-momentum does not contradict to the existence of the process
\begin{equation}\label{SD020}
k_1+k_2+\dots+k_{2n+1} \to \mathcal{K}
\end{equation}
However, the matrix element of this process vanishes exactly due to the structure of the initial Heisenberg-Euler Lagrangian, described by Eqs.~(\ref{eq:Lagrangian}) and~(\ref{eq:HELagr}). Taking into account vanishing photon mass in vacuum, the conservation law corresponding to the reaction~(\ref{SD020}) can be fulfilled only when $\pmb{k}_i\parallel \pmb{\mathcal{K}}$ that leads to vanishing of the the product of the invariant bi-linear forms $f$ and $g$~(\ref{N010})-(\ref{N020}), entering Eq.~(\ref{P096}). This exact cancellation was not taken into account in Ref.~\cite{KS15}, when  the relevant integrals evaluated approximately. It leads to considerable overestimation of effectiveness of the 3rd order harmonic generation. As a result, the lowest order process giving contribution to the process of HOH generation is described by Eqs.~(\ref{P090}) and~(\ref{P160})-(\ref{P080}), corresponding to the process~(\ref{SD010}), as considered in Sec.~\ref{mon}. 

However, when we have two more or less weakly focused colliding beams of finite widths, then the matrix element of the process~(\ref{SD010}), expressed in terms of maximal field strength, contains additional small multiplier $\sim  W^{-2n}$ as considered in Sec.~\ref{comp23}. Here $ W$ characterizes the ratio of the typical waist of the beams to their wavelengths. This small factor does not act for the $4\pi$-dipole $in$-coming wave, considered in Sec.~\ref{3d_dipole}. However, this small factor relevant for colliding Gaussian laser beams can be suppressed in the higher orders of the process relative to dimensionless value of typical field in the focus, $E/E_S$. For example the matrix elements of the non-linear electric current of the process:
$$
k_1+k_2+\dots+k_{3n+1}
$$
\begin{equation}\label{SD030}
\to k_{3n+2}+\dots+k_{4n+1}+\mathcal{K}
\end{equation}
will not contain at all the small parameter $ W^{-2}$ in its leading order. A particular case of the process~(\ref{SD030}) was considered in Ref.~\cite{Sa21}. We showed in Sec.~\ref{comp23} that if the 2nd beam power  is less than a critical value, defined in Eq.~(\ref{CA120}),
then the process~(\ref{SD010}) is more probable than the process~(\ref{SD030}), regardless of the focusing degree, $ W^{-1}$, while it is small.  However, the processes~(\ref{SD030}), considered actually in Ref.~\cite{Sa21}, dominates over~(\ref{SD010}), when the power of the 2nd incident laser beam becomes higher than the critical value under the same other limitations.

\begin{acknowledgments}
This research was funded by the project ``Advanced Research using High Intensity Laser produced Photons and Particles'' (ADONIS) (CZ.02.1.01/0.0/0.0/16019/0000789) from European Regional Development Fund (ERDF)  and  by Ministry of Education Youth and Sports of Czech Republic (MEYS CR) grant number CZ.02.1.01/0.0/0.0/16\_019/0000778.

\end{acknowledgments}


\appendix

\section{Photon states with given $\omega$, $\pmb{j}^2$, $j_z$ and parity}
\label{EMj}

Consider at first the states $Mj$. They have the parity +1 for odd $j$ and the parity -1 for  even $j$, where $j$ is total angular moment of the state $j=1,2,\dots$. It is convenient to introduce the following scalar function.
\begin{equation}\label{C010}
\psi(r,\theta,\varphi,t)=j_j(\omega r)\, Y_{j,m}(\theta,\varphi)e^{-i\omega t}\, .
\end{equation}
Here, $j_j(z)$ are the spherical Bessel functions, and $Y_{\ell m}$ are usual scalar spherical functions. They obey the following relationships:
$$
\partial_\varphi Y_{j,m}=imY_{j,m}\,, 
$$
and
$$
\left[\frac{1}{\sin\theta}\partial_\theta\sin\theta\, \partial_\theta 
-\frac{1}{\sin^2\theta}\, \partial_\varphi^2\right] Y_{j,m}=-j(j+1)\, Y_{j,m}\, .
$$
Thus~\cite{DLMF},
$$
Y_{j,m}\propto\frac{e^{im\varphi}}{\sin^m\theta}\frac{d^{\ell-m}}{(d \cos\theta)^{\ell-m}}\sin^{2\ell}\theta\, ,
$$
and
$$
j_j(z)=z^j\left(-\frac{1}{z}\frac{d}{dz}\right)^j\frac{\sin z}{z}\, .
$$

The EM field corresponding to the state $Mj$ is described by the expressions:
\begin{equation}\label{C020}
(A_r,\,A_\theta,\, A_\varphi) \propto 
 \left(0,\,\frac{-im}{\sin\theta}\psi,\, \partial_\theta\psi\right) \, .
\end{equation}
\begin{equation}\label{C030}
(E_r,\,E_\theta,\, E_\varphi) = 
 \left(0,\,\frac{-im}{\sin\theta}\psi,\, \partial_\theta\psi\right)
\end{equation}
\begin{equation}\label{C040}
B_r=-\frac{i}{\omega}\frac{j(j+1)}{r}\psi\,,
\quad
B_\theta=\frac{i}{\omega r} \partial_r(r\partial_\theta\psi)
\end{equation}
\begin{equation}\label{C050}
B_\varphi=-\frac{m}{\omega r \sin\theta}\partial_r(r\psi)
\end{equation}
The vector potential~(\ref{C020}) is written assuming the transverse gauge, $\nabla\cdot\pmb{A}=0$. Any choice of the gauge fixing is allowed for the calculation of $\bar{\pmb{j}}_{\Omega;\, \pmb{n}\Omega}$ because of its gauge invariance.

The states $Ej$ can be obtained from the states $Mj$ by applying the symmetry $\pmb{E}\leftrightarrow -\pmb{B}$. The parity changes its sign under this transformation.
These states $Mj$ and $Ej$ are parameterized by the following 4 `quantum numbers': $\omega$, $j$, $m=j_z\, (=0,\pm1,\dots ,\pm j)$ and the parity $(\pm 1)$. The parity of the states $Ej$ is equal to $(-1)^j$ and to $(-1)^{j+1}$ for the $Mj$ states. This representation of the set of photon states was considered in the textbook~\cite{Pan05} and originated from Ref.~\cite{He36}.

\begin{widetext}

\section{The case 3D dipole $in$-coming field. Details of the derivations.}
\label{barHK}

\subsection{General expressions}

We derive in this Appendix the explicit expressions for $\bar{H}_n(\Theta)$ and $\bar{K}_n(\Theta)$, introduced in Sec.~\ref{3d_dipole} for the $4\pi$-dipole incoming field. They are defined in Eqs.~(\ref{T140a}) and~(\ref{T150a}) and enter the definition~(\ref{I210})  of  $h_n$. The latter value enters the expression~(\ref{I200}) for the total power of the HOH. We skip temporally in the calculations of this section the multiplier $E_0$ in the expressions~(\ref{T010})-(\ref{T050}).

For $f\left(\tilde{\pmb{A}}^{in},\tilde{\pmb{A}}^{in}\right)$ and $f\left(\tilde{\pmb{A}}^{in},\tilde{\pmb{A}}^{in\, *}\right)$ we have
\begin{equation}\label{T060}
f\left(\tilde{\pmb{A}}^{in},\tilde{\pmb{A}}^{in}\right)=\frac{1}{2}\biggl[j_1^2(\omega r)\sin^2\theta+\frac{1}{\omega^2 r^2} \left[\partial_r\left(r j_1(\omega r)\right)\right]^2\sin^2\theta
+\frac{4}{\omega^2 r^2}j_1^2(\omega r)\cos^2\theta\biggr]\, e^{-2i\omega t}
\end{equation}
\begin{equation}\label{T070}
f\left(\tilde{\pmb{A}}^{in},\tilde{\pmb{A}}^{in\, *}\right)=\frac{1}{2}\left[j_1^2(\omega r)\sin^2\theta-\frac{1}{\omega^2 r^2} \left[\partial_r\left(r j_1(\omega r)\right)\right]^2\sin^2\theta
-\frac{4}{\omega^2 r^2}j_1^2(\omega r)\cos^2\theta\right]
\end{equation}

We use the following expressions for the spatial factors in the out-going plane wave: $\bar{\pmb{\mathcal{A}}}^{out}=\pmb{e}_{\pmb{n}} e^{i\Omega \pmb{n}\pmb{x}}$. Here, $\pmb{n}=(\cos\Theta,0,\sin\Theta)$, $\pmb{e}_{\pmb{n}}=(-\sin\Theta,0,\cos\Theta)$ in the Cartesian coordinate system $(x,y,z)$, associated with the spherical coordinate system, and $\Omega=(2n+1)\omega$. For the time-dependent electric and magnetic components of  the probe out-going wave we have:
$$
\tilde{\pmb{\mathcal{E}}}^{out}=i\Omega\left[-\pmb{e}_x\sin\Theta+\pmb{e}_z\cos\Theta\right] \exp[-i\Omega t
+i\Omega r \left(\cos\Theta\sin\theta\cos\varphi+\sin\Theta\cos\theta\right)]\,,
$$
$$
\tilde{\pmb{\mathcal{B}}}^{out}=-i\Omega\,\pmb{e}_y\exp[-i\Omega t
+i\Omega r \left(\cos\Theta\sin\theta\cos\varphi+\sin\Theta\cos\theta\right)]\,,
$$
whereas the other polarization, when $\tilde{\pmb{\mathcal{E}}}^{out}\parallel\pmb{e}_y$ for the same $\pmb{n}$, will not give finally any contribution to $\pmb{j}_{\perp\, \Omega;\, \pmb{n}\Omega}$. Due to azimuthal symmetry of the in-coming wave we may set $\pmb{e}_y\cdot\pmb{n}=0$. Further we obtain:
$$
f\left(\tilde{\pmb{A}}^{in},\tilde{\pmb{\mathcal{A}}}^{out\, *}\right)=
-\frac{i(2n+1)\omega}{2}\exp\left[2ni\omega t-(2n+1)i\omega r
\left(\cos\Theta\sin\theta\cos\varphi+\sin\Theta\cos\theta\right)\right]
$$
$$
\times\biggl\{-j_1(\omega r)\left[\sin\theta\cos\varphi-\frac{2i}{\omega r}\left(\cos\Theta\cos^2\theta+\sin\Theta\sin\theta\cos\theta\cos\varphi\right)\right]
$$
\begin{equation}\label{T080}
+\frac{i}{\omega r}\left(\partial_r(rj_1(\omega r))\right)\left[\cos\Theta\sin^2\theta-\sin\Theta\cos\theta\sin\theta\cos\varphi\right]\biggl\}
\end{equation}
$$
f\left(\tilde{\pmb{A}}^{in\, *},\tilde{\pmb{\mathcal{A}}}^{out\, *}\right)=
-\frac{i(2n+1)\omega}{2}\exp\left[(2n+2)i\omega t-(2n+1)i\omega r
\left(\cos\Theta\sin\theta\cos\varphi+\sin\Theta\cos\theta\right)\right]
$$
$$
\times\biggl\{-j_1(\omega r)\left[\sin\theta\cos\varphi+\frac{2i}{\omega r}\left(\cos\Theta\cos^2\theta+\sin\Theta\sin\theta\cos\theta\cos\varphi\right)\right]
$$
\begin{equation}\label{T090}
-\frac{i}{\omega r}\left(\partial_r(rj_1(\omega r))\right)\left[\cos\Theta\sin^2\theta-\sin\Theta\cos\theta\sin\theta\cos\varphi\right]\biggr\}
\end{equation}
To calculate these expressions we used the following representation of the in-coming field in the Cartesian coordinate system:
$$
\pmb{e}_r=\pmb{e}_x \sin\theta\cos\varphi+\pmb{e}_y \sin\theta\sin\varphi+\pmb{e}_z \cos\theta\, ,
$$
$$
\pmb{e}_\theta=\pmb{e}_x \cos\theta\cos\varphi+\pmb{e}_y \cos\theta\sin\varphi-\pmb{e}_z \sin\theta\, ,
$$
$$
\pmb{e}_\varphi=-\pmb{e}_x \sin\varphi+\pmb{e}_y \cos\varphi\, ,
$$
$$
\tilde{E}_x^{in}=\tilde{E}_r^{in}\sin\theta\cos\varphi+\tilde{E}_\theta^{in} \cos\theta\cos\varphi\,,
$$
$$
\tilde{E}_y^{in}=\tilde{E}_r^{in}\sin\theta\sin\varphi+\tilde{E}_\theta^{in} \cos\theta\sin\varphi\,,
$$
$$
\tilde{E}_z^{in}=\tilde{E}_r^{in}\cos\theta-\tilde{E}_\theta^{in} \sin\theta\,,
$$
$$
\tilde{B}_x^{in}=-\tilde{B}_\varphi^{in}\sin\varphi\,,
\qquad 
\tilde{B}_y^{in}=\tilde{B}_\varphi^{in}\cos\varphi\,.
$$

The multipliers~(\ref{T060}) and~(\ref{T070}) do not depend on $\varphi$ and remain the same under the transform $\theta\to\pi-\theta$. The multipliers enter the definitions~(\ref{P170}) and~(\ref{P180}) of $H _n$ and $K_n$, respectively.
Hence, calculating the integrals in the definitions, we can  replace $f\left(\tilde{\pmb{A}}^{in},\tilde{\pmb{\mathcal{A}}}^{out\, *}\right)$ and $f\left(\tilde{\pmb{A}}^{in\, *},\tilde{\pmb{\mathcal{A}}}^{out\, *}\right)$ by their mean values $\overline{f\left(\tilde{\pmb{A}}^{in},\tilde{\pmb{\mathcal{A}}}^{out\, *}\right)}$ and $\overline{f\left(\tilde{\pmb{A}}^{in\, *},\tilde{\pmb{\mathcal{A}}}^{out\, *}\right)}$, obtained by averaging over $\varphi$ and over the substitution $\theta\leftrightarrow\pi-\theta$:
$$
\overline{f\left(\tilde{\pmb{A}}^{in},\tilde{\pmb{\mathcal{A}}}^{out\, *}\right)}=\frac{(2n+1)\omega}{2}\, \exp\left(2ni\omega t\right)
$$
$$
\times\Biggl\{\cos\left[(2n+1)\omega r
\sin\Theta\cos\theta\right]\biggl[j_1(\omega r)\left(J_1\sin\theta+\frac{2}{\omega r}J_0\cos\Theta\cos^2\theta \right)
+\frac{1}{\omega r}\left(\partial_r(rj_1(\omega r))\right)J_0\cos\Theta\sin^2\theta\biggr]
$$
\begin{equation}\label{T100}
-\frac{1}{\omega r}\sin\left[(2n+1)\omega r
\sin\Theta\cos\theta\right]\,\left[ j_1(\omega r)- \omega r j_1^\prime(\omega r)\right] \, J_1\sin\Theta\sin\theta\cos\theta\Biggr\}
\end{equation}
$$
\overline{f\left(\tilde{\pmb{A}}^{in\, *},\tilde{\pmb{\mathcal{A}}}^{out\, *}\right)}=\frac{(2n+1)\omega}{2}\, \exp\left((2n+2)i\omega t\right)
$$
$$
\times\Biggl\{\cos\left[(2n+1)\omega r
\sin\Theta\cos\theta\right]\biggl[j_1(\omega r)\left(J_1\sin\theta-\frac{2}{\omega r}J_0\cos\Theta\cos^2\theta \right)
-\frac{1}{\omega r}\left(\partial_r(rj_1(\omega r))\right)J_0\cos\Theta\sin^2\theta\biggr]
$$
\begin{equation}\label{T110}
+\frac{1}{\omega r}\sin\left[(2n+1)\omega r
\sin\Theta\cos\theta\right]\,\left[j_1(\omega r)-\omega r j_1^\prime(\omega r)\right] \, J_1\sin\Theta\sin\theta\cos\theta\Biggr\}
\end{equation}
We use here the following shortcut in two latter equations: $J_m=J_m((2n+1)\,\omega r \cos\Theta\sin\theta)$ $(m=0,\, 1)$. 

The intensity of the harmonic with the number $2n+1$ can be expressed through the matrix elements~(\ref{P170}) and~(\ref{P180}):
$$
H_n
=-\int f^{n}\left(\tilde{\pmb{A}}^{in},\tilde{\pmb{A}}^{in}\right)\,f\left(\tilde{\pmb{A}}^{in},\tilde{\pmb{A}}^{in\, *}\right) \, f\left(\tilde{\pmb{A}}^{in},\tilde{\pmb{\mathcal{A}}}^{out\, *}\right)\, d\pmb{x}
$$
\begin{equation}\label{T120}
=-2\pi\int\limits_0^\infty r^2\, dr \int\limits_0^\pi\sin\theta\, d\theta\,  f^{n}\left(\tilde{\pmb{A}}^{in},\tilde{\pmb{A}}^{in}\right)\,f\left(\tilde{\pmb{A}}^{in},\tilde{\pmb{A}}^{in\, *}\right) \, \overline{f\left(\tilde{\pmb{A}}^{in},\tilde{\pmb{\mathcal{A}}}^{out\, *}\right)}\,,
\end{equation}
\begin{equation}\label{T130}
K_n
=-\int f^{n+1}\left(\tilde{\pmb{A}}^{in},\tilde{\pmb{A}}^{in}\right)\, f\left(\tilde{\pmb{A}}^{in\, *},\tilde{\pmb{\mathcal{A}}}^{out\, *}\right)\, d\pmb{x} =-2\pi\int\limits_0^\infty r^2\, dr \int\limits_0^\pi\sin\theta\, d\theta\, f^{n+1}\left(\tilde{\pmb{A}}^{in},\tilde{\pmb{A}}^{in}\right)\, \overline{f\left(\tilde{\pmb{A}}^{in\, *},\tilde{\pmb{\mathcal{A}}}^{out\, *}\right)}\, .
\end{equation}
All factors in the integrands of these expressions are defined explicitly in Eqs.~(\ref{T060}), (\ref{T070}), (\ref{T100}) and~(\ref{T110}). Explicit expressions for spatial integrals $\bar{H}_n$ and $\bar{K}_n$, defined in Eqs.~(\ref{T140a}) and~(\ref{T150a}), are presented in the next subsection of this appendix.

Eqs.~(\ref{T060}), (\ref{T070}), (\ref{T100}), (\ref{T110}), (\ref{T120}) and~(\ref{T130}) indicate that imaginary parts of the `matrix elements' $H_n$ and $K_n$ are equal to 0. This conclusion depends on the choice of the phase multiplier in the expressions~(\ref{T040}) and~(\ref{T050}).

\subsection{Explicit expressions for the spatial integrals in Eqs.~(\ref{T120}) and~(\ref{T130})}

Combining the expressions~~(\ref{T120}) and~(\ref{T130}) and expressions~(\ref{T060}),  (\ref{T070}), (\ref{T100}) and~(\ref{T110}) for the invariant $f$,  using the definitions~(\ref{T140a}) and~(\ref{T150a}), and recovering the amplitude multiplier $E_0$ in Eqs.~(\ref{T010})-(\ref{T050}), we obtain the following explicit expressions for $\bar{H}_n(\Theta)$ and $\bar{K}_n(\Theta)$:
$$
\bar{H}_n(\Theta)=-2\int\limits_0^\infty \rho^2\, d\rho\int\limits_0^{\pi/2} \sin\theta\, d\theta\, \left[j_1^2(\rho)\sin^2\theta+\frac{1}{\rho^2} \left[\partial_\rho\left(\rho j_1(\rho)\right)\right]^2\sin^2\theta
+\frac{4}{\rho^2}j_1^2(\rho)\cos^2\theta\right]^n
$$
$$
\times\left[j_1^2(\rho)\sin^2\theta-\frac{1}{\rho^2} \left[\partial_\rho\left(\rho j_1(\rho)\right)\right]^2\sin^2\theta
-\frac{4}{\rho^2}j_1^2(\rho)\cos^2\theta\right]
$$
$$
\times\Biggl\{\cos\left[(2n+1)\rho
\sin\Theta\cos\theta\right]\biggl[j_1(\rho)\left(J_1\sin\theta+\frac{2}{\rho}J_0\cos\Theta\cos^2\theta \right)
+\frac{1}{\rho}\left(\partial_\rho(\rho j_1(\rho))\right)J_0\cos\Theta\sin^2\theta\biggr]
$$
\begin{equation}\label{T142}
-\frac{1}{\rho}\sin\left[(2n+1)\rho
\sin\Theta\cos\theta\right]\,\left[ j_1(\rho)-\rho j_1^\prime(\rho)\right] \, J_1\sin\Theta\sin\theta\cos\theta\Biggr\}\, ,
\end{equation}
where $J_m=J_m((2n+1)\,\rho \cos\Theta\sin\theta)$ $(m=0,\, 1)$.
$$
\bar{K}_n(\Theta)=-2\int\limits_0^\infty \rho^2\, d\rho\int\limits_0^{\pi/2} \sin\theta\, d\theta\, \left[j_1^2(\rho)\sin^2\theta+\frac{1}{\rho^2} \left[\partial_\rho\left(\rho j_1(\rho)\right)\right]^2\sin^2\theta
+\frac{4}{\rho^2}j_1^2(\rho)\cos^2\theta\right]^{n+1}
$$
$$
\times\Biggl\{\cos\left[(2n+1)\rho
\sin\Theta\cos\theta\right]\biggl[j_1(\rho)\left(J_1\sin\theta-\frac{2}{\rho}J_0\cos\Theta\cos^2\theta \right)
-\frac{1}{\rho}\left(\partial_\rho(\rho j_1(\rho))\right)J_0\cos\Theta\sin^2\theta\biggr]
$$
\begin{equation}\label{T160}
+\frac{1}{\rho}\sin\left[(2n+1)\rho
\sin\Theta\cos\theta\right]\,\left[ j_1(\rho)-\rho j_1^\prime(\rho)\right] \, J_1\sin\Theta\sin\theta\cos\theta\Biggr\}\, .
\end{equation}
These integrals as functions of $\Theta$ are calculated numerically. The results are used in the main body of the paper to plot Figs.~\ref{fig3d1}-\ref{fig3d7} ant to calculate $h_n$ defined in Eq.~(\ref{I220}). The latter coefficient enters our main results~(\ref{I200}) and~(\ref{CC102}) for the $4\pi$-dipole incoming wave and is presented in Tab.~\ref{gh_3d}. We have checked that our numerical procedure gives that $|H(n,0,0)|\ll H_n$. This situation corresponds to the exact equality: $H(n,0,0)=0$. Our calculations show also that $|K_n|$ considerably less than $|H_n|$ for $n=3$ and 4. Moreover, $H_n$ are positive. It means that Fig.~\ref{fig3d7} together with Tab.~\ref{gh_3d} gives almost complete information about $H_n(\Theta)$ at $n=3$ and 4.

\section{The case of two crossing Gaussian beams. Details of the derivations.}
\label{3dGauss}

\subsection{One Gaussian beam in paraxial approximations with subleading corrections}

We consider in this section the exact expression~(\ref{T420}) for the in-coming field in the limit $ W\to\infty$. It can be considered as an exact model for the 3d Gaussian beam.

We are able to make the following substitution in the exponent at $ W\to\infty$, whereas $\omega\sqrt{y^2+z^2}\sim\mathcal{O}( W)$ and $\omega|x|\sim\mathcal{O}( W^2)$:
$$
i\omega x\left(1-\frac{\zeta^2+\varphi^2}{2}\right)+i\omega y\varphi+i\omega z \zeta
-\frac{ W^2}{2}\left(\varphi^2+\zeta^2\right)\, ,
$$
Residual terms can be moved to the pre-exponent. Thus,
\begin{equation}\label{T430}
\pmb{e}_y\cdot\bar{\pmb{A}}\left(\pmb{x}\right)=\frac{ W^2}{2\pi}
\int d\varphi\,  d\zeta\, Q_y(\pmb{x},\zeta,\varphi)\,\exp\left[i\omega x\left(1-\frac{\zeta^2+\varphi^2}{2}\right)+i\omega y\varphi+i\omega z \zeta
-\frac{ W^2}{2}\left(\varphi^2+\zeta^2\right)\right]\, ,
\end{equation}
where the polynomial $Q(\pmb{x},\zeta,\varphi)$ is equal to
$$
Q_y(\pmb{x},\zeta,\varphi)=1-\zeta ^2-\frac{\varphi ^2}{2} 
+\frac{ W^2}{6}\left(\zeta ^4+3\zeta ^2 \varphi ^2 +\varphi ^4\right)  
+\frac{i x \omega}{24}  \left(\zeta ^4  +6 \zeta ^2 \varphi ^2  + \varphi ^4 \right)
-\frac{i y \omega\varphi}{6}\left(3\zeta ^2 +\varphi ^2\right)-\frac{1}{6} i z\omega  \zeta ^3+\dots\, .
$$
The dots mean residual terms of the order of $\mathcal{O}( W^{-3})$ at $\omega\sqrt{y^2+z^2}\sim\mathcal{O}( W)$ and $\omega|x|\sim\mathcal{O}( W^2)$ and $ W\to\infty$.
\begin{equation}\label{T440}
\pmb{e}_x\cdot\bar{\pmb{A}}\left(\pmb{x}\right)=\frac{ W^2}{2\pi}
\int d\varphi\,  d\zeta\, Q_x(\pmb{x},\zeta,\varphi)\,\exp\left[i\omega x\left(1-\frac{\zeta^2+\varphi^2}{2}\right)+i\omega y\varphi+i\omega z \zeta
-\frac{ W^2}{2}\left(\varphi^2+\zeta^2\right)\right]\, ,
\end{equation}
where
$$
Q_x(\pmb{x},\zeta,\varphi)=-\varphi+\dots\, .
$$
\begin{equation}\label{T444}
\pmb{e}_z\cdot\bar{\pmb{A}}\left(\pmb{x}\right)=0\,.
\end{equation}
These equations give an explicit expression for $\bar{\pmb{A}}_\omega\left(\pmb{x}\right)$ at $ W\to\infty$, whereas $\omega\sqrt{y^2+z^2}\sim\mathcal{O}( W)$ and $\omega|x|\sim\mathcal{O}(W^2)$.
\begin{equation}\label{T452}
\pmb{e}_y\cdot\bar{\pmb{A}}\left(\pmb{x}\right)=\frac{W^2}{W^2+i\omega x}\exp\left[i\omega x-\frac{\omega^2(y^2+z^2)}{2\left(W^2+i\omega x\right)}\right]\,\left[1+R_1+\dots\right]\, ,
\end{equation}
where
$$
R_1=-\frac{3W^2+3i\omega x -\omega^2 (2z^2+y^2)}{2\left(W^2+i\omega x\right)^2}
$$
$$
+W^2
\frac{6\left(W^2+i\omega x\right)^2-6\left(W^2+i\omega x\right)\omega^2(y^2+z^2)+\omega^4(y^4+z^4)+3\left(W^2+i\omega x-\omega^2y^2\right)\left(W^2+i\omega x-\omega^2z^2\right)}{6\left(W^2+i\omega x\right)^4}
$$
$$
+ix\omega
\frac{6\left(W^2+i\omega x\right)^2-6\left(W^2+i\omega x\right)\omega^2(y^2+z^2)+\omega^4(y^4+z^4)+6\left(W^2+i\omega x-\omega^2y^2\right)\left(W^2+i\omega x-\omega^2z^2\right)}{24\left(W^2+i\omega x\right)^4}
$$
\begin{equation}\label{T450}
+\frac{\omega^2(3 y^2+zy+z^2)\left(W^2+i\omega x-\omega^2z^2\right)}{6\left(W^2+i\omega x\right)^3}\,.
\end{equation}
\begin{equation}\label{T460}
\pmb{e}_x\cdot\bar{\pmb{A}}\left(\pmb{x}\right)=\frac{W^2}{W^2+i\omega x}\exp\left[i\omega x-\frac{\omega^2(y^2+z^2)}{2\left(W^2+i\omega x\right)}\right]\, \left[R_2+\dots\right]\,,
\end{equation}
where
\begin{equation}\label{T462}
R_2=-\frac{iy\omega}{2\left(W^2+i\omega x\right)}\,.
\end{equation}

\end{widetext}
 
Another polarization of the beam can be obtained by the substitution $\pmb{E}\leftrightarrow -\pmb{B}$, or setting $\pmb{n}_{\pmb{k}}\times\pmb{a}_{\pmb{k}}$ instead of $\pmb{a}_{\pmb{k}}$ that is defined by Eq.~(\ref{T424}).

It is interesting to evaluate typical value of the invariant $g$ for the field, defined by Eq.~(\ref{T424}), when $W\gg1$. The electric and magnetic fields are defined in this case by the expressions~(\ref{T444})-(\ref{T460}). As a result we have:
\begin{equation}\label{T470}
g(\tilde{\pmb{A}},\tilde{\pmb{A}})=\omega e^{-2i\omega t}
\left(\bar{A}_{x}\partial_z\bar{A}_{y}
-\bar{A}_{y}\partial_z\bar{A}_{x}\right).
\end{equation}
The subscript indexes $x$, $y$ and $z$ in this expression mean $x$-, $y$- and $z$-components of the vector $\bar{\pmb{A}}$, respectively.
Anti-symmetric entering of the components of the vector potential in the expression~(\ref{T470}) notes that we can differentiate only the last, 3rd multipliers in Eq.~(\ref{T450}) and~(\ref{T460}). As a result, we obtain for this state:
\begin{equation}\label{T472}
g(\tilde{\pmb{A}},\tilde{\pmb{A}})\sim\frac{\omega^2A_0^2}{W^4}e^{2i\omega t}\, ,
\end{equation}
when $W\to\infty$, while $\omega\sqrt{y^2+z^2}\sim\mathcal{O}(W)$ and $\omega|x|\sim\mathcal{O}(W^2)$. The analogous statement is valid for $g(\tilde{\pmb{A}},\tilde{\pmb{A}}^{*})$:
\begin{equation}\label{T473}
g(\tilde{\pmb{A}},\tilde{\pmb{A}}^{*})\sim\frac{\omega^2 |A_0|^2}{W^4}\, .
\end{equation}
 Meanwhile, asymptotic behavior of the invariant $f$ for this sate looks like:
\begin{equation}\label{T480}
|f|\sim\frac{\omega^2\left|A_0\right|^2}{W^2}\, .
\end{equation}
See the next subsection.

\subsection{Calculation of $f\left(\tilde{\pmb{A}}_1,\tilde{\pmb{A}}_1\right)$ at $|\pmb{x}|\omega \sim W$ in the limit} $W\to\infty$
\label{a1a1}

To obtain leading terms of $H_n$ and $K_n$ at $W\to\infty$ we calculate here a leading term of $f\left(\tilde{\pmb{A}}_1,\tilde{\pmb{A}}_1\right)$ at the same limit for the space domain $|\pmb{x}|\omega \sim W$. To do this we represent the expressions~(\ref{T450}) and~(\ref{T460}) in the following form:
\begin{equation}\label{CC210}
\left(\bar{\pmb{A}}_1\right)_y\approx e^{i x} G \left[1+R_1\right]\, ,
\end{equation}
and
\begin{equation}\label{CC220}
\left(\bar{\pmb{A}}_1\right)_x\approx e^{i x} G R_2\, .
\end{equation}
Here
$$
G=\frac{W^2}{W^2+ix}\exp\left[-\frac{(y^2+z^2)}{2\left(W^2+i x\right)}\right]\, ,
$$
$$
R_2=-\frac{iy}{2(W^2+i x)}\, ,
$$
and $R_1\sim{\cal O}\left(W^{-2}\right)$ is determined by the rest of the expression~(\ref{T450}). We set here temporally $A_0=\omega=1$

Then 
$$
\bar{B}_z=\partial_x A_y-\partial_y A_x
$$
$$
\approx ie^{i x} G \left[1+R_1\right]-ie^{ix}GW^{-2}+\frac{i}{2}e^{ix}G\frac{(y^2+z^2)}{W^4}
$$
\begin{equation}\label{CC224}
\approx i e^{i x} G\left[1+R_1 -i\frac{2W^2-y^2-z^2}{2W^4}\right]\, ,
\end{equation}
\begin{equation}\label{CC230}
\bar{B}_x=-\partial_z A_y\approx +\frac{z}{W^2}e^{i x} G\, ,
\end{equation}
\begin{equation}\label{CC234}
\bar{B}_y=\partial_z A_x\approx 0\, ,
\end{equation}
\begin{equation}\label{CC238}
\bar{E}_x=ie^{i x} G R_2\, ,\quad \bar{E_z}=0\,,
\end{equation}
\begin{equation}\label{CC240}
\bar{E}_y\approx i e^{i x} G \left[1+R_1\right]\, .
\end{equation}

Thus, calculation of $f\left(\tilde{\pmb{A}}_1,\tilde{\pmb{A}}_1\right)$ at this limit gives
$$
f\left(\tilde{\pmb{A}}_1,\tilde{\pmb{A}}_1\right)=\left(-\tilde{E}_x^2-\tilde{E}_y^2+\tilde{B}_x^2+\tilde{B}_y^2\right)/2\approx
$$
$$
\frac{1}{2}e^{2i(x-t)}G^2\Biggl[R_2^2+1+2R_1-1-2R_1
$$
$$
+2i\frac{2W^2-y^2-z^2}{2W^4}-\frac{z^2}{W^4}\Biggr]
$$
$$
\approx\frac{1}{2}e^{2i(x-t)}G^2\Biggl[2i\frac{2W^2-y^2-z^2}{2W^4}-\frac{z^2+y^2/4}{W^4}\Biggr]
$$
$$
\approx\frac{e^{2i(x-t)}}{2W^2}\exp\left[-\frac{y^2+z^2}{W^2}\right]
$$
\begin{equation}\label{CC250}
\times\left[2i -(1+i)\frac{y^2+z^2}{W^2}+\frac{3y^2}{4W^2}\right]\, .
\end{equation}
Recovering $\omega$ and $A_0$ we obtain:
$$
f\left(\tilde{\pmb{A}}_1,\tilde{\pmb{A}}_1\right)=A_0^2\frac{e^{2i\omega (x-t)}}{2w_e^2}\exp\left[-\frac{(y^2+z^2)}{w_e^2}\right]
$$
\begin{equation}\label{CC260}
\times\left[2i -(1+i)\frac{y^2+z^2}{w_e^2}+\frac{3y^2}{4w_e^2}\dots\right]\, .
\end{equation}

\subsection{Calculation of $f(\tilde{\pmb{A}}_1,\tilde{\pmb{A}}^{out\, *})$ at $|\pmb{x}|\omega =\mathcal{O}(W)$, $\pmb{n}\cdot\pmb{e}_x=1-\mathcal{O}\left(W^{-2}\right)$ and $W\to\infty$}
\label{a1a_out}

To obtain leading terms of $H_n$ and $K_n$ at $W\to\infty$ we calculate here a leading term of $f\left(\tilde{\pmb{A}}_1,\tilde{\pmb{\mathcal{A}}}^{out\, *}\right)$ at $W\to\infty$ in the space domain $|\pmb{x}|\omega =\mathcal{O}(W)$. We set here
\begin{equation}\label{CC270}
\bar{\pmb{\mathcal{A}}}^{out}=\pmb{e}_{\pmb{n}} e^{i\Omega\pmb{n}\pmb{x}}\, ,
\end{equation}
where
\begin{equation}\label{CC272}
\pmb{n}=(\cos\Theta,\,\sin\Theta\, \cos{\phi},\, \sin\Theta\, \sin{\phi})
\end{equation}
is the direction of the HOH propagation, and
$$
\pmb{e}_{\pmb{n}}= \pmb{n}\times\pmb{e}_{\pmb{n}}^\prime 
$$
$$
\pmb{e}_{\pmb{n}}^\prime
=\frac{\pmb{n}\times\pmb{e}_y}{|\pmb{n}\times\pmb{e}_y|}=\frac{(-\sin\Theta\, \sin\phi,\, 0,\, \cos\Theta)}{\sqrt{\cos^2\Theta+\sin^2\Theta\sin^2\phi}}\, .
$$
$$
\pmb{e}_{\pmb{n}}
$$
$$
=\frac{(\sin\Theta\cos\Theta\cos\phi, \sin^2\Theta\cos^2\phi-1, \sin^2\Theta\cos\phi\sin\phi)}{\sqrt{\cos^2\Theta+\sin^2\Theta\sin^2\phi}} .
$$
We will see that the matrix elements are not exponentially small only when $\Theta\sim\mathcal{O}\left(W^{-1}\right)$. In this case, the rate emission of the harmonic with perpendicular polarization,
$\pmb{e}_{\pmb{n}}^\prime$, will contain an additional small factor $\sim W^{-2}$ in comparison with the $\pmb{e}_{\pmb{n}}$ polarization.

For the fields component of this wave we have:
$$
\bar{\pmb{\mathcal{B}}}^{out}=i\Omega\, \pmb{e}_{\pmb{n}}\times \pmb{n}\, e^{i\Omega\pmb{n}\pmb{x}}=i\Omega\, \pmb{e}_{\pmb{n}}^\prime\, e^{i\Omega\pmb{n}\pmb{x}}\, .
$$
$$
\bar{\pmb{\mathcal{E}}}^{out}=i\Omega\, \pmb{e}_{\pmb{n}}\, e^{i\Omega\pmb{n}\pmb{x}}\, .
$$
$$
\left(\tilde{\pmb{\mathcal{B}}}^{out}\right)_x\approx -i\Omega\, \Theta \sin\phi \, e^{i\Omega(\pmb{n}\pmb{x}-t)}\,,
$$
$$
\left(\tilde{\pmb{\mathcal{B}}}^{out}\right)_y=0\,,
$$
$$
\left(\tilde{\pmb{\mathcal{B}}}^{out}\right)_z\approx i\Omega\,  e^{i\Omega(\pmb{n}\pmb{x}-t)}\,,
$$
$$
\left(\tilde{\pmb{\mathcal{E}}}^{out}\right)_x\approx i\Omega\, \Theta\cos\phi\, e^{i\Omega(\pmb{n}\pmb{x}-t)}\, ,
$$
$$
\left(\tilde{\pmb{\mathcal{E}}}^{out}\right)_y\approx i\Omega\, (\Theta^2\cos^2\phi-1)\, e^{i\Omega(\pmb{n}\pmb{x}-t)}\, ,
$$
$$
\left(\tilde{\pmb{\mathcal{E}}}^{out}\right)_z\approx 0\, .
$$
Thus, we obtain
$$
f\left(\tilde{\pmb{A}}_1,\tilde{\pmb{\mathcal{A}}}^{out\, *}\right)=\left(\tilde{\pmb{B}}_1\tilde{\pmb{\cal B}}^{out\, *}-\tilde{\pmb{E}}_1\tilde{\pmb{\cal E}}^{out\, *}\right)/2\approx
$$
$$
\approx i\Omega\Theta\sin\phi\frac{z}{2W^2}Ge^{i(x-t)+i\Omega(t-\pmb{n}\pmb{x})}
$$
$$
+\frac{\Omega }{2}G\left[1+R_1 -i\frac{2W^2-y^2-z^2}{2W^4}\right]e^{i (x-t)+i\Omega(t-\pmb{n}\pmb{x})}
$$
$$
+\frac{\Omega}{2} \Theta \cos\phi \, G R_2e^{i (x-t)+i\Omega(t-\pmb{n}\pmb{x})}
$$
$$
+ \frac{\Omega}{2}   G \left[1+R_1\right](\Theta^2\cos^2\phi-1) e^{i (x-t)+i\Omega(t-\pmb{n}\pmb{x})}
$$
$$
\approx \frac{\Omega}{2} G\,  e^{i (x-t)+i\Omega(t-\pmb{n}\pmb{x})}\Biggl\{\frac{z\Theta\sin\phi}{W^2}-\frac{iy\Theta\cos\phi}{2W^2}
$$
$$
+\left[1+R_1 -i\frac{2W^2-y^2-z^2}{2W^4}
-1-R_1+\Theta^2\cos^2\phi\right]\Biggr\}
$$
$$
=\frac{\Omega}{2} \exp\left[-\frac{(y^2+z^2)}{2W^2}\right]\,  e^{i (x-t)+i\Omega(t-\pmb{n}\pmb{x})}
$$
$$
\times\biggl[\frac{z\Theta\sin\phi}{W^2}-\frac{iy\Theta\cos\phi}{2W^2}
$$
\begin{equation}\label{CC280}
 -i\frac{2W^2-y^2-z^2}{2W^4}
+\Theta^2\cos^2\phi\biggr]\, .
\end{equation}

Recovering $\omega$ and $A_0$ we obtain:
$$
f\left(\tilde{\pmb{A}}_1,\tilde{\pmb{\mathcal{A}}}^{out\, *}\right)=\frac{A_0\Omega\omega}{2W^2} \exp\left[-\frac{(y^2+z^2)}{2w_e^2}\right]\, 
$$
$$ 
\times e^{i\omega (x-t)+i\Omega(t-\pmb{n}\pmb{x})}\biggl[\left(\frac{z\sin\phi}{w_e}-\frac{iy\cos\phi}{2w_e}\right)\Theta W
$$
\begin{equation}\label{CC290}
 -i\frac{2w_e^2-y^2-z^2}{2w_e^2}
+(\Theta W)^2\cos^2\phi\biggr]\, .
\end{equation}

It is interesting that the 1st post-paraxial correction $R_1$ is canceled in the expressions for both $f\left(\tilde{\pmb{A}}_1,\tilde{\pmb{A}}_1\right)$ and  $f(\tilde{\pmb{A}}_1,\tilde{\pmb{\cal A}}^{out\, *})$, whereas the analogous correction $R_2$, entering as $R_2^2$, gives contributions to the both expressions.

\begin{widetext}

\subsection{Final expressions for the matrix elements $H_n^{(2)}$, $H_n^{(4)}$ and $K_n$ for two crossing Gaussian beams}
\label{H_K}

Substituting lowest orders final expressions for $f\left(\tilde{\pmb{A}}_1,\tilde{\pmb{A}}_1\right)$ [Eq.~(\ref{CC260})], $f\left(\tilde{\pmb{A}}_1,\tilde{\pmb{\mathcal{A}}}^{out\, *}\right)$ [Eq.~(\ref{CC290})], $f(\tilde{\pmb{A}}_1,\tilde{\pmb{A}}_2)$ [Eq.~(\ref{CA010})], $f(\tilde{\pmb{A}}_1,\tilde{\pmb{A}}_2^*)$  [Eq.~(\ref{CA020})], $f(\tilde{\pmb{A}}_2,\tilde{\pmb{A}}^{out\, *})$  [Eq.~(\ref{CA030})], and $f(\tilde{\pmb{A}}_2^*,\tilde{\pmb{A}}^{out\, *})$ [Eq.~(\ref{CA040})] into the expressions~(\ref{T540}), (\ref{T560}) and~(\ref{T562}) for $H_n^{(2)}$, $H_n^{(4)}$ and $K_n$, respectively, we obtain the following final expressions for the matrix elements:
$$
H_n^{(2)}=-2n\int A_{1\,,0}^{2(n-1)}\frac{e^{2i(n-1)\omega (x-t)}}{2^{n-1}w_e^{2(n-1)}}\exp\left[-\frac{(n-1)(y^2+z^2)}{w_e^2}\right]
\left[2i -(1+i)\frac{y^2+z^2}{w_e^2}+\frac{3y^2}{4 w_e^2}\right]^{n-1}
$$
$$
\times\frac{A_{1\,,0}\Omega\omega}{2W^2} \exp\left[-\frac{(y^2+z^2)}{2w_e^2}\right]\,
e^{i\omega (x-t)+i\Omega(t-\pmb{n}\pmb{x})}\biggl[\left(\frac{z\sin\phi}{w_e}+\frac{iy\cos\phi}{w_e}\right)\Theta W
 -i\frac{2w_e^2-y^2-z^2}{2w_e^2}
+(\Theta W)^2\cos^2\phi\biggr]
$$
$$
\times (-1)\frac{A_{1\,,0}A_{2\,,0}\omega^2}{2}e^{i\omega(x+z-2t)}\exp\left(-\frac{x^2+2y^2+z^2}{2w_e^2}\right)
\frac{A_{1\,,0}A_{2\,,0}^*\omega^2}{2}e^{i\omega(x-z)}\exp\left(-\frac{x^2+2y^2+z^2}{2w_e^2}\right)
\, dx\,dy\,dz
$$
$$
=\frac{(2n+1)n}{2^{n+2}}\, \frac{A_{1\,0}^{2n+1}|A_{2\,,0}|^2\omega^{2n+4}}{W^{2n}}
\int e^{i (2n+1)\omega(x-\pmb{n}\pmb{x})}\, \exp\left(-\frac{2x^2 + (2n+3)y^2+(2n+1)z^2}{2w_e^2}\right)
$$
\begin{equation}\label{CC300}
\times
\left[2i -(1+i)\frac{y^2+z^2}{w_e^2}+\frac{3y^2}{4 w_e^2}\right]^{n-1}
\biggl[\left(\frac{z\sin\phi}{w_0}+\frac{iy\cos\phi}{w_0}\right)\Theta W
 -i\frac{2w_e^2-y^2-z^2}{2w_e^2}
+(\Theta W)^2\cos^2\phi\biggr]
\, dx\,dy\,dz\, ,
\end{equation}
$$
H_n^{(4)}=-\int A_{1\,,0}^{2n}\frac{e^{2in\omega (x-t)}}{2^n w_e^{2n}}\exp\left[-\frac{n(y^2+z^2)}{w_e^2}\right]
\left[2i -(1+i)\frac{y^2+z^2}{w_e^2}+\frac{3y^2}{4 w_e^2}\right]^{n}
$$
$$
\times(-1)\frac{A_{2\,,0}\omega\Omega}{2}e^{i\Omega(t-\pmb{n}\pmb{x})+i\omega(z-t)}\exp\left(-\frac{x^2+y^2}{2w_e^2}\right)
\frac{A_{1\,,0}A_{2\,,0}^*\omega^2}{2}e^{i\omega(x-z)}\exp\left(-\frac{x^2+2y^2+z^2}{2w_e^2}\right)
\, dx\,dy\,dz
$$
$$
=\frac{2n+1}{2^{n+2}}\,\frac{A_{1\,,0}^{2n+1}|A_{2\,,0}|^2\omega^{2n+4}}{W^{2n}}
\int e^{i\omega (2n+1)(x-\pmb{n}\pmb{x})}
\exp\left(-\frac{2x^2+(2n+3)y^2+(2n+1)z^2}{2w_e^2}\right)
$$
\begin{equation}\label{CC310}
\times \left[2i -(1+i)\frac{y^2+z^2}{w_e^2}+\frac{3y^2}{4 w_e^2}\right]^{n}
\, dx\,dy\,dz\, ,
\end{equation}
$$
K_n=-2(n+1)\int A_{1\,,0}^{2n}\frac{e^{2in\omega (x-t)}}{2^nw_e^{2n}}\exp\left[-\frac{n(y^2+z^2)}{w_e^2}\right]
\left[2i -(1+i)\frac{y^2+z^2}{w_e^2}+\frac{3y^2}{4 w_e^2}\right]^{n}
$$
$$
\times (-1)\frac{A_{1\,,0}A_{2\,,0}\omega^2}{2}e^{i\omega(x+z-2t)}\exp\left(-\frac{x^2+2y^2+z^2}{2w_e^2}\right)
\frac{A_{2\,, 0}^*\omega\Omega}{2}e^{i\Omega(t-\pmb{n}\pmb{x})+i\omega(t-z)}\exp\left(-\frac{x^2+y^2}{2w_e^2}\right)
\, dx\,dy\,dz
$$
$$
=\frac{(n+1)(2n+1)}{2^{n+2}}\,\frac{A_{1\,,0}^{2n+1}|A_{2\,,0}|^2\omega^{2n+4}}{W^{2n}}
\int e^{i\omega (2n+1)(x-\pmb{n}\pmb{x})}
\exp\left(-\frac{2x^2+(2n+3)y^2+(2n+1)z^2}{2w_e^2}\right)
$$
\begin{equation}\label{CC320}
\times 
\left[2i -(1+i)\frac{y^2+z^2}{w_e^2}+\frac{3y^2}{4 w_e^2}\right]^{n}
\, dx\,dy\,dz\, .
\end{equation}

\subsection{Final expressions for the combined matrix element 
entering the expression (\ref{P180a}) for HOHs intensities produced by  crossing Gaussian beams}
\label{calK}

The coefficients $H_n^{(2)}$, $H_n^{(4)}$ and $K_n$ enter the expression~(\ref{P180a}) for the harmonics intensity only in the combination:
\begin{equation}\label{CC330}
\mathcal{K}_n=(2n+2)H_n^{(2)}+(2n+2)H_n^{(4)}+K_n\, .
\end{equation}
Using Eqs.~(\ref{CC300}), (\ref{CC310}) and~(\ref{CC320}), we obtain:
$$
\mathcal{K}_n=\frac{(2n+1)(n+1)}{2^{n+2}}\, \frac{A_{1\,,0}^{2n+1}|A_{2\,,0}|^2\omega^{2n+4}}{W^{2n}}
\int
\left[2i -(1+i)\frac{y^2+z^2}{w_e^2}+\frac{3y^2}{4 w_e^2}\right]^{n-1}
$$
$$
\times\left\{2n\biggl[\left(\frac{z\sin\phi}{w_e}+\frac{iy\cos\phi}{w_e}\right)\Theta W
 -i\frac{2w_e^2-y^2-z^2}{2w_e^2}
+(\Theta W)^2\cos^2\phi\biggr]+3\left[2i -(1+i)\frac{y^2+z^2}{w_e^2}+\frac{3y^2}{4 w_e^2}\right]\right\}
$$
\begin{equation}\label{CC340}
 \times\exp\left(-i(2n+1)\omega \Theta y\cos\phi - i(2n+1)\omega\Theta z\sin\phi-\frac{2x^2+(2n+3)y^2+(2n+1)z^2}{2w_e^2}\right)
\, dx\,dy\,dz\, .
\end{equation}
Introducing the scaled angle of scattering,
\begin{equation}\label{CC350}
\tilde{\Theta}=W\Theta
\end{equation}
we obtain
$$
\mathcal{K}_n(\Theta,\phi)=\mathcal{K}_n(\tilde{\Theta}/W,\phi)=\frac{(2n+1)(n+1)}{2^{n+2}}\, \frac{A_{1\,,0}^{2n+1}|A_{2\,,0}|^2\omega^{2n+1}}{W^{2n-3}}\int
\left[2i -(1+i)(\upsilon^2+\zeta^2)+\frac{3}{4}\upsilon^2\right]^{n-1}
$$
$$
\times
\left\{2n\biggl[\left(\zeta\sin\phi+i\upsilon\cos\phi\right)\tilde{\Theta}
 -i\left( 1-\frac{\upsilon^2+\zeta^2}{2}\right)
+\tilde{\Theta}^2\cos^2\phi\biggr]+3\left[2i -(1+i)(\upsilon^2+\zeta^2)+\frac{3}{4}\upsilon^2\right]\right\}
$$
$$
 \exp\left(-i(2n+1)\tilde{\Theta} \upsilon\cos\phi - i(2n+1)\tilde{\Theta} \zeta\sin\phi-\chi^2-\frac{(2n+3)\upsilon^2+(2n+1)\zeta^2}{2}\right)
\, d\chi\,d\upsilon\,d\zeta
$$
$$
=\frac{(2n+1)(n+1)\sqrt{\pi}}{2^{n+2}}\, \frac{A_0^{2n+1}|A_0|^2\omega^{2n+1}}{W^{2n-3}}
\int
\left[2i -(1+i)(\upsilon^2+\zeta^2)+\frac{3}{4}\upsilon^2\right]^{n-1}
$$
$$
\times\left\{2n\biggl[\left(\zeta\sin\phi+i\upsilon\cos\phi\right)\tilde{\Theta}
 -i\left( 1-\frac{\upsilon^2+\zeta^2}{2}\right)
+\tilde{\Theta}^2\cos^2\phi\biggr]+3\left[2i -(1+i)(\upsilon^2+\zeta^2)+\frac{3}{4}\upsilon^2\right]\right\}
$$
$$
 \exp\left(-i(2n+1)\tilde{\Theta} \upsilon\cos\phi - i(2n+1)\tilde{\Theta} \zeta\sin\phi-\frac{(2n+3)\upsilon^2+(2n+1)\zeta^2}{2}\right)
\, d\upsilon\,d\zeta
$$
$$
=\frac{(2n+1)(n+1)\sqrt{\pi}}{2^{n+2}}\, \frac{A_{1\,,0}^{2n+1}|A_{2\,,0}|^2\omega^{2n+1}}{W^{2n-3}}
\int
\left[2i -(1+i)(\upsilon^2+\zeta^2)+\frac{3}{4}\upsilon^2\right]^{n-1}
$$
$$
\times\left\{2n\biggl[\zeta\tilde{\Theta}_\zeta+i\upsilon\tilde{\Theta}_\upsilon
 -i\left( 1-\frac{\upsilon^2+\zeta^2}{2}\right)
+\tilde{\Theta}_\upsilon^2\biggr]+3\left[2i -(1+i)(\upsilon^2+\zeta^2)+\frac{3}{4}\upsilon^2\right]\right\}
$$
\begin{equation}\label{CC360}
 \exp\left(-i(2n+1)\tilde{\Theta}_\upsilon \upsilon - i(2n+1)\tilde{\Theta}_\zeta \zeta-\frac{(2n+3)\upsilon^2+(2n+1)\zeta^2}{2}\right)
\, d\upsilon\,d\zeta\, ,
\end{equation}
where
$\tilde{\Theta}_\upsilon=\tilde{\Theta}\cos\phi$ and $\tilde{\Theta}_\zeta=\tilde{\Theta}\sin\phi$.

We define:
\begin{equation}\label{CC370}
\mathcal{K}_n=\frac{(2n+1)(n+1)\sqrt{\pi}}{2^{n+2}}\, \frac{A_{1\,,0}^{2n+1}|A_{2\,,0}|^2\omega^{2n+1}}{W^{2n-3}}\, \tilde{\mathcal{K}}_n\left(\tilde{\Theta}_\upsilon,\tilde{\Theta}_\zeta\right)\, ,
\end{equation}
where
$$
\tilde{\mathcal{K}}_n\left(\tilde{\Theta}_\upsilon,\tilde{\Theta}_\zeta\right)=\int
\left[2i -(1+i)(\upsilon^2+\zeta^2)+\frac{3}{4}\upsilon^2\right]^{n-1}
$$
$$
\times\left\{2n\biggl[\zeta\tilde{\Theta}_\zeta+i\upsilon\tilde{\Theta}_\upsilon
 -i\left( 1-\frac{\upsilon^2+\zeta^2}{2}\right)
+\tilde{\Theta}_\upsilon^2\biggr]+3\left[2i -(1+i)(\upsilon^2+\zeta^2)+\frac{3}{4}\upsilon^2\right]\right\}
$$
\begin{equation}\label{CC380}
 \times\exp\left(-i(2n+1)\tilde{\Theta}_\upsilon \upsilon - i(2n+1)\tilde{\Theta}_\zeta \zeta-\frac{(2n+3)\upsilon^2+(2n+1)\zeta^2}{2}\right)
\, d\upsilon\,d\zeta\, .
\end{equation}
This integral can be easily calculated analytically for any integer $n$, because it can be presented a sum of iterated integrals that are composed from the single integrals of the form
$$
\int\limits_{-\infty}^\infty u^\ell e^{bu-au^2}\, du
$$
with $a>0$ and integer $\ell$.
Nevertheless, the general expression for any $n$ is too complicated to be presented here. We used Wolfram MATHEMATICA v.13.2 for obtaining the analytic representations of the integral for $n=1$, 2, 3 and 4. Thus, we have
\begin{equation}\label{CC390}
\tilde{\mathcal{K}}_1=\frac{\pi}{150\sqrt{15}}\left[-350+880 i +(1041+216 i)\tilde{\Theta}_\upsilon^2+900\tilde{\Theta}_\zeta^2\right]\, \exp\left(-\frac{9\tilde{\Theta}_\upsilon^2+15\tilde{\Theta}_\zeta^2}{10}\right)\, .
\end{equation}
$$
\tilde{\mathcal{K}}_2=\frac{\pi}{480200\sqrt{35}}\,\Bigl[-2 355 479 - 1 154 048 i+(70 + 280 i) (31 495 + 19 724)\tilde{\Theta}_\upsilon^2- (1 248 520 - 4 829 440 i)\tilde{\Theta}_\zeta^2
$$
\begin{equation}\label{CC392}
+(636 875 + 3 610 000 i) \tilde{\Theta}_\upsilon^4 +5 762 400\tilde{\Theta}_\zeta^4 +(70 + 280 i)(35 700 - 20 720 i)\tilde{\Theta}_\upsilon^2\tilde{\Theta}_\zeta^2\Bigr] \,
\exp\left(-\frac{25\tilde{\Theta}_\upsilon^2+35\tilde{\Theta}_\zeta^2}{14}\right)\, .
\end{equation}
$$
\tilde{K}_3=\frac{\pi }{5833096416 \sqrt{7}} \exp\left(-\frac{49\tilde{\Theta}_\upsilon^2+56\tilde{\Theta}_\zeta^2}{18}  \right)
$$
$$ 
\times\biggl\{(-14836127145+7912601144 i)\tilde{\Theta}_\upsilon^6+(64827+259308 i)\tilde{\Theta}_\upsilon^4 \left[(238812+210000 i)\tilde{\Theta}_\zeta^2
+(-51959+282172 i)\right]
$$
$$
+5103\tilde{\Theta}_\upsilon^2 \left[(4840416+24554880 i)\tilde{\Theta}_\zeta^4
-(35020104-20075328 i)\tilde{\Theta}_\zeta^2+(-22187571-4451624 i)\right]
$$
$$
+2187 \Bigl[(21337344+10668672 i)\tilde{\Theta}_\zeta^6-(11219040-37594368 i)\tilde{\Theta}_\zeta^4
$$
\begin{equation}\label{CC420}
-(12581100-12845952 i)\tilde{\Theta}_\zeta^2+(1830801+904208 i)\Bigr]\biggr\}\, .
\end{equation}

$$
\tilde{K}_4=-\frac{\pi }{180020303134848 \sqrt{11}} \exp\left(-\frac{81\tilde{\Theta}_\upsilon^2+99\tilde{\Theta}_\zeta^2}{22} \right)
$$
$$ \biggl\{(417425043611583+422468096116896 i)\tilde{\Theta}_\upsilon^8+(8418025440+15783797700 i)\tilde{\Theta}_\upsilon^6
$$
$$
\times \left[(139392-171468 i)\tilde{\Theta}_\zeta^2+(193888+25199 i)\right]
$$
$$
-(4763286+19053144 i)\tilde{\Theta}_\upsilon^4 \Bigl[(93148704+469611648 i)\tilde{\Theta}_\zeta^4
$$
$$
-(669506904-456131808 i)\tilde{\Theta}_\zeta^2+(-446502741-63683164 i)\Bigr]
$$
$$
+47916\tilde{\Theta}_\upsilon^2 \Bigl[(79569692928-176816939904 i)\tilde{\Theta}_\zeta^6+(451570176288+76512408192 i)\tilde{\Theta}_\zeta^4
$$
$$
+(240361127028+354312569952 i)\tilde{\Theta}_\zeta^2+(-8177943955+187838676760 i)\Bigr]
$$
$$
-14641 \Bigl[(98365031424+196730062848 i)\tilde{\Theta}_\zeta^8-(463342353408-361830965760 i)\tilde{\Theta}_\zeta^6
$$
\begin{equation}\label{CC410}
-(575406203328-1304709120 i)\tilde{\Theta}_\zeta^4-(346926531600+93483168768 i)\tilde{\Theta}_\zeta^2+(-95165128391-1776051584 i)\Bigr]\biggr\}\, .
\end{equation}
These four expressions were used for the analytic calculations of $h_n$, entering Eq.~(\ref{CA050}), and defined in Eq.~(\ref{CC400a}), as well as for plotting Fig.~\ref{figC1}.

\end{widetext}



\begin{thebibliography}{99}
\bibitem{Lund2} We consider here only the cases, when $n\geq 1$. Nevertheless, the case $n=0$, considered earlier in Ref.~\cite{Lu06}, belongs to the same way of reasoning.
\end{thebibliography}
\end{document}